\documentclass[aps,superscriptaddress,twocolumn,twoside,floatfix,pra,a4paper]{revtex4-2}
\usepackage{times}
\usepackage{amsmath,amssymb,amsthm,mathtools,thmtools,nameref}
\usepackage{comment}
\usepackage{enumerate}
\usepackage{float}
\usepackage{xcolor}
\usepackage[colorlinks=true,linkcolor=blue,citecolor=magenta,urlcolor=blue]{hyperref}
\usepackage[framemethod=default]{mdframed}
\usepackage{csvsimple}
\usepackage[mode=tex]{standalone}
\usepackage[capitalize]{cleveref}

\usepackage{pgfplots}
\pgfplotsset{compat=1.18} 
\usepackage{tikz}

\newcommand{\tr}{\text{Tr}}
\newcommand{\ket}[1]{|#1\rangle}
\newcommand{\bra}[1]{\langle#1|}

\newcommand{\Rho}[1]{\rho_{b^{(#1)}}}
\newcommand{\A}[1]{A_{b^{(#1)}}}
\newcommand{\Support}[1]{P_{b^{(#1)}}}
\newcommand{\W}[1]{W_{b^{(#1)}}}
\newcommand{\K}[1]{K_{b^{(#1)}}}
\newcommand{\Q}[1]{Q_{b^{(#1)}}}
\newcommand{\B}[1]{B_{b^{(#1)}}}
\newcommand{\V}[1]{V_{b^{(#1)}}}
\newcommand{\D}[1]{D_{b^{(#1)}}}

\begin{document}
\title{High-fidelity, multi-qubit generalized measurements with dynamic circuits
}
\author{Petr Ivashkov}
\
\affiliation{Department of Information Technology and Electrical Engineering, ETH Z\"urich, Z\"urich, Switzerland}
\author{Gideon Uchehara}
\affiliation{Department of Electrical and Computer Engineering, University of British Columbia, Vancouver, Canada}
\author{Liang Jiang}
\affiliation{Pritzker School of Molecular Engineering, University of Chicago, Chicago, IL, USA}
\author{Derek~S.~Wang}
\affiliation{IBM Quantum, IBM T.J. Watson Research Center, Yorktown Heights, NY, USA}
\author{Alireza Seif}
\affiliation{IBM Quantum, IBM T.J. Watson Research Center, Yorktown Heights, NY, USA}

\begin{abstract}
Generalized measurements, also called positive operator-valued measures (POVMs), can offer advantages over projective measurements in various quantum information tasks. Here, we realize a generalized measurement of one and two superconducting qubits with high fidelity and in a single experimental setting. To do so, we propose a hybrid method, the ``Naimark-terminated binary tree," based on a hybridization of Naimark's dilation and binary tree techniques that leverages emerging hardware capabilities for mid-circuit measurements and feed-forward control. Furthermore, we showcase a highly effective use of approximate compiling to enhance POVM fidelity in noisy conditions. We argue that our hybrid method scales better toward larger system sizes than its constituent methods and demonstrate its advantage by performing detector tomography of symmetric, informationally complete POVM (SIC-POVM). Detector fidelity is further improved through a composite error mitigation strategy that incorporates twirling and a newly devised conditional readout error mitigation. Looking forward, we expect improvements in approximate compilation and hardware noise for dynamic circuits to enable generalized measurements of larger multi-qubit POVMs on superconducting qubits.
\vspace{1cm}
\end{abstract}

\maketitle
\twocolumngrid

\section{Introduction}

Measuring information accurately and efficiently from inherently probabilistic systems is a central challenge of quantum physics. Projective, or von Neumann, measurements are often used in experiments because of their comparatively simple realization on many quantum computing platforms \cite{Nielsen2012}. At the same time, they can be suboptimal in such tasks as quantum state discrimination~\cite{Bergou2010DiscriminationOfQuantumStates, Braunstein1994}, where no projective measurement can unambiguously tell two non-orthogonal states apart with a single shot. Generalized measurements or positive operator-valued measures (POVMs) define the most general framework for quantum measurements, including projective measurement as a special case. 
Among various broad areas~\cite{Gisin2007, Briegel2009, Tth2014}, POVMs allow for unambiguous state discrimination~\cite{Bergou2010DiscriminationOfQuantumStates, Bae2015StateDiscrimination}, optimal state tomography~\cite{Scott2006,Renes2004,Haah2016}, entanglement detection~\cite{Shang2018EntanglementDetection}, Bell's inequalities~\cite{Bene2010BellInequality}, quantum machine learning algorithms~\cite{Yun2022POVMQML}, and improved observable estimates in variational quantum algorithms~\cite{Perez2021LearningtoMeasure}. Therefore, it is crucial to have a deterministic protocol, which can be realized in a single experimental setting to implement a general POVM~\cite{comment_probabilistic_methods}.

Despite the high utility of POVMs, realizing them on superconducting quantum systems is challenging. In principle, any POVM can be realized by a projective measurement in an extended Hilbert space through Naimark's dilation~\cite{Naimark1976, Peres1990NeumarkTheorem}. Recent efforts involved embedding the system in the qudit space of superconducting qubits~\cite{Fischer2022} and trapped ions~\cite{Stricker2022}. However, this requires efficient discrimination of qudit states, which adds a level of experimental complexity and suffers from readout errors. Alternatively, a POVM may be realized by coupling the system to a number of auxiliary qubits that scales with the size of the POVM~\cite{Yordanov2019, pinto2023povmstateprep}. Such implementations, however, raise concerns about circuit complexity and, hence, scalability to multi-qubit systems. For instance, the large number of auxiliary qubits required for the Naimark's dilation may not be readily available or directly connected to the qubits one intends to measure. Overall, Naimark's dilation encounters practical implementation issues due to the complex unitary operations required in the extended Hilbert space. 

A promising alternative is a binary search, which employs only a single auxiliary qubit to realize general multi-qubit POVMs~\cite{Andersson2008, Shen2017}. It involves a sequence of conditional two-outcome POVMs, requiring cutting-edge hardware capabilities such as mid-circuit measurements and feed-forward control~\cite{Corcoles2021Dynamic, Baumer2023LongRange} comprising ``dynamic" or ``adaptive" circuits. This scheme has been recently demonstrated in a specialized experiment with a single microwave cavity coupled to a transmon qubit~\cite{Cai2021TransmonCavity}. However, extending the implementation of this scheme to multi-qubit programmable quantum processors requires circuits with potentially large depths and many feed-forward operations that limit its fidelity. 

To address the limitation imposed on POVM fidelity by circuit noise, we propose a novel method for single-setting POVMs on multi-qubit systems using dynamic circuits. Furthermore, we introduce innovative use of approximate compiling to implement measurements, enhancing POVM fidelity under noisy conditions~\cite{Madden2022}. In Section~\ref{sec:hybridscheme}, we present a hybridization of Naimark's dilation with the binary search---an approach that we call ``Naimark-terminated binary tree". Our hybrid method results in shorter-depth circuits and scales better toward larger systems. In Section~\ref{sec:experiment}, we implement all three methods on an IBM quantum device and demonstrate the advantage of our hybrid approach by performing detector tomography of symmetric, informationally complete POVM (SIC-POVM)~\cite{Renes2004, Scott2006}. Using our hybrid approach with a composite error mitigation strategy including twirling and newly devised conditional readout error mitigation (CREM), we improve the fidelity of two qubit SIC-POVM to $70.4\pm0.1\%$ from $52\%$ and $40\%$ of bare Naimark and binary tree approaches, respectively. Our code and data are available on GitHub~\footnote{The repository contains code for performing POVMs on multi-qubit systems, including implementations of Naimark's dilation, binary search, and the hybrid approach. It also includes data and notebooks for generating plots equivalent to the plots in the paper. \url{https://github.com/petr-ivashkov/dynamic-circuit-povms}.}.

\section{POVM} \label{sec:povm}
Formally, a POVM is a set $F = \{F_i\}$ of $M$ positive semi-definite Hermitian operators, called POVM elements. Each element $F_i$ corresponds to a measurement outcome $i$ with probability $\mathrm{P(i)} = \tr(F_i\rho)$, where $\rho$ is the state of the system. POVM elements must satisfy the completeness relation $\sum_{i=1}^{M}F_i = I$ to have a normalized probability distribution. Unlike projective measurements, POVM elements are not necessarily orthogonal. We restrict our attention to POVMs whose elements are linearly independent rank-one operators $F_i = \ket{\psi_i}\bra{\psi_i}$ where $\ket{\psi_i}$ is not necessarily normalized. Any higher-rank POVMs can be obtained by relabeling and mixing the outcomes of rank-one POVMs with a maximum of $d^2$ elements, where $d$ is the dimension of the system~\cite{Haapasalo2011}. Note that a POVM only defines the measurement statistics $\mathrm{P(i)}$ but not the post-measurement state of the system, which depends on how the POVM is realized. In fact, we disregard the post-measurement state in applications like observable estimation or quantum state tomography (see Appendix~\ref{sec:state_tomography}), where the system is measured only once at the end of the experiment. Therefore, this paper focuses only on measurement statistics. This allows us to measure both the system and the auxiliary qubits directly using Naimark's dilation and achieve a higher fidelity than binary search at the cost of destroying the post-measurement state.

\section{Naimark-terminated binary tree}\label{sec:hybridscheme}
In this section, we propose a novel combination of two previously known methods for general POVMs, binary search and Naimark’s dilation. This new hybrid scheme, the Naimark-terminated binary tree, is more efficient than its constituent methods. In a nutshell, we perform the binary search by repeatedly dividing the set of POVM elements in half to narrow down the search range. When the number of remaining POVM elements corresponds to the dimension of the compound system and auxiliary Hilbert space, we interrupt the binary search and apply Naimark's dilation. 

We begin with the binary search. As originally detailed by Andersson and Oi~\cite{Andersson2008}, binary search can realize general POVM with only a single auxiliary qubit. It, therefore, reduces the complexity of manipulating an extended system and saves the quantum memory space when $M$ is large. We adopt the notation from Shen et al.~\cite{Shen2017} to briefly outline the key steps below; see Appendix~\ref{sec:details_binary_tree} for the full treatment.

To construct the binary search tree, we begin by padding our set of POVM elements with zero operators until $M$ is the nearest power of two. In the first step, we split the original POVM into two sets of $\frac{M}{2}$ elements, for example as
\begin{equation}
    B_0 := \sum_{i=1}^{M/2}F_i  \hspace{15pt}\text{and}\hspace{15pt} B_1 := \sum_{i=M/2+1}^{M}F_i.
\end{equation}
The ordering of non-zero $F_i$ may be arbitrary and corresponds to relabelling the measurement outcomes. The set $\{B_0, B_1\}$ constitutes a valid POVM, which can be realized via an indirect measurement of the system using a single auxiliary qubit. Specifically, we measure the auxiliary qubit after it has interacted with the system via a suitable coupling unitary, effectively implementing a completely positive map~\cite{Nielsen2012}. The corresponding Kraus operators must satisfy the isometry condition $A_0^{\dagger}A_0 + A_1^{\dagger}A_1 = I$. Since the POVM $\{B_0, B_1\}$ is complete, we can always find suitable Kraus operators by taking the square root of the corresponding POVM elements: $A_0 = \sqrt{B_0}$ and $A_1 = \sqrt{B_1}$. Finally, we construct the coupling unitary $U^{B}$ by stacking together the two binary Kraus operators $A_0$ and $A_1$ and completing the remaining matrix elements: 
\begin{equation}
    U^{B} = \begin{pmatrix}
        A_0 & *\hspace{8pt}	\\
        A_1 & *\hspace{8pt} \\
    \end{pmatrix} \in \mathbb{U}(2d).
    \label{eq:stinespring_matrix}
\end{equation}
The unitary operation is followed by the measurement of the auxiliary qubit in the computational basis. Finally, the auxiliary qubit is reset in the $\ket{0}$ state.

\begin{figure*}[!tb]
    \centering\includegraphics{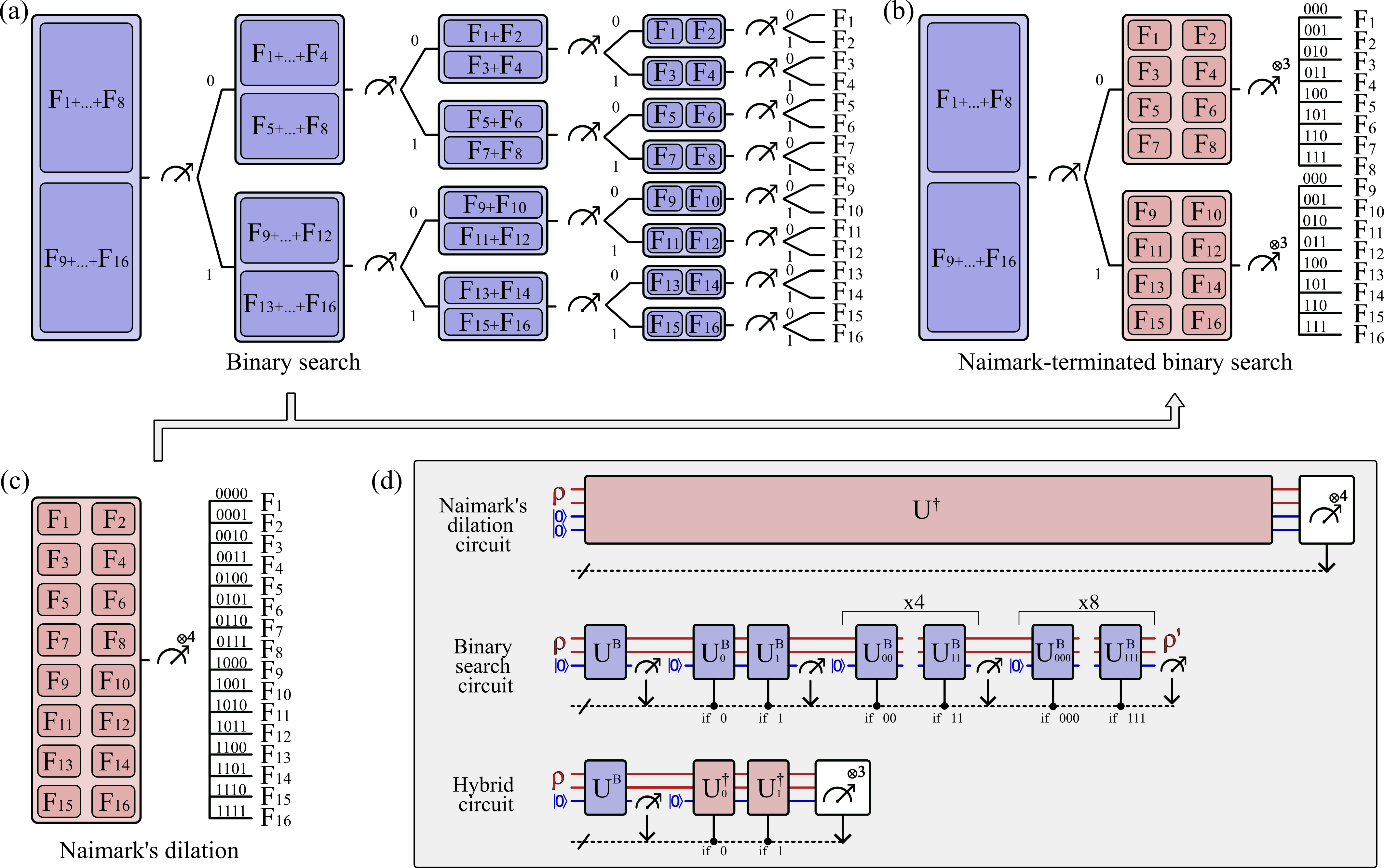}
    \caption{Schematic of a two-qubit POVM with $M$=16 POVM elements realized through \textbf{(a)} binary tree (blue), \textbf{(b)} hybrid scheme (blue and red), and \textbf{(c)} Naimark's dilation (red). Branching occurs after the measurement conditional on the outcome, with the trajectory defined by a bitstring of previous outcomes. Each scheme terminates with a bitstring of length four corresponding to one of 16 POVM elements. The number of qubits $n$ and the number of POVM elements $M$ can be arbitrary. Grey panel \textbf{(d)} shows quantum circuit implementations for methods \textbf{(a)}-\textbf{(c)}, ordered from bottom to top. System qubits are in red, and auxiliary qubits are in blue. Binary search tree and hybrid circuit include mid-circuit measurements and unitaries conditional on the classical register (dotted line). Longer unitary blocks correspond to the higher circuit depth.}
    \label{fig:two_qubit_schematic}
\end{figure*}

After this first partial filtering, we again split the remaining $\frac{M}{2}$ POVM elements in each branch into $\frac{M}{4}$ elements. This time, however, to implement the next step of partial filtering by measuring a binary POVM, we have to account for the post-measurement state by modifying the Kraus operators for the subsequent steps $l\ge2$ as follows:
\begin{equation}
    \A{l} = \K{l} \K{l-1}^{-1},
    \label{eq:kraus_operator_hybrid}
\end{equation}
where we use a binary string $b^{(l)}$ of length $l$ to denote the sequence of measurement outcomes leading to the current branch in the binary search tree, \textit{e.g.} $b^{(1)} \in \{0,1\}$. Here, $\K{l} = \sqrt{\sum_{i=a}^{b}F_i}$ is obtained by aggregating the POVM elements located in the last level of the branch that starts from $b^{(l)}$ with indices ranging from $a$ to $b$. 
Importantly, at every filtering step, we partition POVM elements such that each branch receives at least $d$ non-zero POVM elements. As a result, the matrix $\K{l}$ is full rank for $l \le m$, making it invertible. It is also worth noting that a unitary transformation $\K{l} \rightarrow \W{l}\K{l}$ with arbitrary unitary $\W{l}$ leaves the measurement statistics invariant and, thus, may have an optimization potential for constructing the coupling unitary in Eq.~\eqref{eq:stinespring_matrix}. As an example, Fig.~\ref{fig:two_qubit_schematic} (b) illustrates how a 16-element POVM on a two-qubit system is realized using the hybrid method. The blue box in Fig.~\ref{fig:two_qubit_schematic} (b) aggregates the corresponding POVM elements.

For POVMs with more than four elements, we continue bisecting the remaining POVM elements within each branch until each branch has at most $2d$ elements in its leaves. Precisely we arrive at this point after $m = \log_2(\frac{M}{2d})$ iterations. For example, a 16-element POVM on two-qubits requires a single iteration of binary search, as shown in Fig.~\ref{fig:two_qubit_schematic} (b). At level $m$ the cumulative Kraus operator is given as
\begin{equation}
    \A{m}^{c} = \A{m} ... \A{2}\A{1} = \K{m},
    \label{eq:kraus_operator_hybrid_m}
\end{equation}
and the conditional post-measurement state of the system is: 
\begin{equation}
    \rho_{b^{(m)}} = \frac{\K{m}\rho \K{m}^{\dagger}}{\tr(\K{m}\rho \K{m}^{\dagger})}.    
    \label{eq:post_measurement_state_hybrid}
\end{equation}
In principle, we could continue dividing the set of POVM elements in half, narrowing down the search range until the target element is found, as shown in Fig. \ref{fig:two_qubit_schematic} (a). This approach would require $\log_2{M} - m$ additional iterations and a modified construction of Kraus operators for subsequent iterations, as detailed in Appendix~\ref{sec:details_binary_tree}.

Instead, we observe that at level $m$, each branch has at most $2d$ elements in its leaves. For example, as illustrated in Fig.~\ref{fig:two_qubit_schematic} (b), each red box aggregates at most eight POVM elements. Therefore, at this point, the single auxiliary qubit suffices to apply Naimark’s dilation. To account for the post-measurement state, we modify the vectors $\ket{\psi_i}$ to $\ket{\tilde{\psi_i}} = \K{m}^{-1\dagger}\ket{\psi_i}$. The new POVM defined by $\tilde{F_i} = \ket{\tilde{\psi_i}}\bra{\tilde{\psi_i}}$ satisfies the completeness relation because
\begin{equation}
    \sum_{i=a}^{b} \ket{\tilde{\psi_i}}\bra{\tilde{\psi_i}} = \K{m}^{-1\dagger} \left(\sum_{i=a}^{b} F_i\right) \K{m}^{-1} = I,
    \label{eq:completeness_relation_hybrid}
\end{equation}
where $a$ and $b$ stand for the first and last indices of aggregated POVM elements at the level $m$ with $b-a+1 = 2d$. Therefore, the modified vectors $\ket{\tilde{\psi_i}}$ can be arranged as column vectors to form a $d\times 2d$ array whose rows are orthonormal $2d$-dimensional vectors. Such an array can always be extended to a $2d\times 2d$ unitary matrix $U$, as detailed in Appendix~\ref{sec:details_naimark_dilation}. By projectively measuring the compound system in the computational basis after the unitary transformation $U^{\dagger}$, as shown in Fig.~\ref{fig:two_qubit_schematic} (b), we obtain the outcomes $\ket{\tilde{\psi_i}^{\mathrm{ext}}}\bra{\tilde{\psi_i}^{\mathrm{ext}}}$, where $\ket{\tilde{\psi_i}^{\mathrm{ext}}}$ are the columns of $U$. The conditional probability of observing the outcome $i$ given that a string of previous measurement outcomes $b^{(m)}$ has been obtained is given by:
\begin{equation}
    \mathrm{P(i | b^{(m)})} = \tr(\ket{\tilde{\psi_i}}\bra{\tilde{\psi_i}} \rho_{b^{(m)}}) = \frac{\tr(F_i\rho)}{\tr(\K{m}\rho \K{m}^{\dagger})},
\end{equation}
which results in the correct measurement statistics $\mathrm{P(i)} = \tr(F_i\rho)$. 

Overall, the hybrid scheme involves $m = \log_2\left(\frac{M}{2d}\right)$ binary search steps and a final level that applies Naimark's dilation. This results in a total of $m+1$ steps where $(n+1)$-qubit unitaries are applied. In contrast, a standard binary search requires $\log_2{M}$ layers of $(n+1)$-qubit unitaries. Therefore, a hybrid circuit has $\log_2{d}$ fewer layers. For instance, the 16-element POVM in Fig.~\ref{fig:two_qubit_schematic} is implemented through the binary search in $\log_2{M} = 4$ steps, each involving a unitary acting on $n+1 = 3$ qubits. In contrast, hybrid involves only a single $(m = \log_2{(\frac{M}{2d})} = 1)$ binary search step and an additional level of conditional Naimark unitaries, thus requiring 2 steps of 3-qubit unitaries. Bare Naimark's dilation realizes the same POVM with a single application of a 4-qubit unitary. Finally, it is worth noting that when $M \le 2d$, the hybrid scheme simplifies to a single application of Naimark's dilation without the need for any binary search steps. For example, a 4-element POVM on a single qubit is most efficiently realized with a single application of Naimark's dilation, as detailed in Appendix~\ref{sec:onequbitschematic}.

\section{Experiment}\label{sec:experiment}
In this section, we compare the three methods by implementing SIC-POVMs on one and two qubits. The choice of SIC-POVMs as our benchmark is motivated by their extensive study in the literature, particularly for their unique tomographic properties~\cite{Renes2004}. Formally, a SIC-POVM consist of $M = d^2$ elements $F_i = \frac{1}{d}\ket{\phi_i}\bra{\phi_i}$ which have equal pairwise overlap with each other $|\bra{\phi_i}\phi_j\rangle|^2 = \frac{1}{d+1}$ for $i\neq j$. This symmetric property makes this class of POVMs particularly important in the context of optimal state tomography~\cite{Sosa-Martinez2017, Stricker2022} and quantum key distribution~\cite{Bent2015SICPOVM, Renes2020SICCompoundsQKD}. However, practical implementation of SIC-POVMs becomes challenging for more than one qubit. 

\subsection{Detector tomography of SIC-POVMs} \label{sec:detector_tomography_main}
In practice, the realized POVM will differ from the ideal SIC-POVM due to noise in the circuit~\cite{IvanovaRohling2023}. To quantify the quality of POVM implementation, we perform detector tomography by preparing an overcomplete set of initial states and reconstructing the realized POVM from measurement statistics, as described in Appendix \ref{sec:detector_tomography_appendix}. To calculate the fidelity, we represent the POVM as a measurement channel~\cite{Wilde2016QuantumInfo, Blumoff2016PreciseMultiqubitMeasurements}:
\begin{equation}
    \mathcal{E}_F(\rho) = \sum_{i=1}^M \tr\left( F_i \rho \right) \ket{i}\bra{i},
    \label{eq:povm_channel}
\end{equation}
where the output state $\mathcal{E}_F(\rho)$ represents the POVM outcome probabilities. The POVM fidelity between the target POVM $F$ and the realized POVM $\tilde{F}$ can be defined as
\begin{equation}
    \mathcal{F}_D(\mathcal{E}_F, \mathcal{E}_{\tilde{F}}) = \mathcal{F}_S(\Lambda_{\mathcal{E}_F}, \Lambda_{\mathcal{E}_{\tilde{F}}}),
\end{equation}
where $\mathcal{F}_S(\rho, \sigma) = \tr\left( \sqrt{\sqrt{\rho}\sigma\sqrt{\rho}}\right)^2$ is the conventional state fidelity. Here, $\Lambda_{\mathcal{E}_F}$ stands for the normalized Choi matrix of the quantum channel:
\begin{equation}
    \Lambda_{\mathcal{E}_F} = \frac{1}{d}\sum_{i,j=1}^d \ket{i}\bra{j} \otimes \mathcal{E}_F(\ket{i}\bra{j}).
    \label{eq:normalized_choi_matrix}
\end{equation}
The POVM fidelity can be reduced by different noise sources in the circuit.
To mitigate the impact of coherent noise, we employ Pauli twirling on CNOTs, taking five twirled instances for each circuit~\cite{knill2004fault,kern2005quantumtwirl,bennet1996purification,geller2013efficienttwirl,Wallman2016Twirling}. Furthermore, the effect of imprecise readout is mitigated using readout error mitigation~\cite{nation2021scalablemitigation,van2022model}. In circuits involving mid-circuit measurement and feed-forward, measurement errors can propagate through the conditional operations and further reduce the fidelity. To address these errors, we extended the standard readout error mitigation to account for error propagation in dynamic circuits. As detailed in Appendix~\ref{sec:crem}, our CREM technique involves taking measurement statistics from additional circuits that provide the necessary information for the deconvolution of noisy outcome probabilities. Finally, we note that the fidelity of a POVM with completely random outcomes is $\frac{1}{2^n}$. Therefore, a non-trivial realization of a POVM should yield a fidelity higher than $\frac{1}{2^n}$. For example, for one qubit, the baseline fidelity is $50 \%$, and for two qubits, it is $25\%$.

\subsection{One-qubit SIC-POVM}\label{sec:onequbitsicpovm}
\begin{figure}[!tb]
    \centering
    \includegraphics{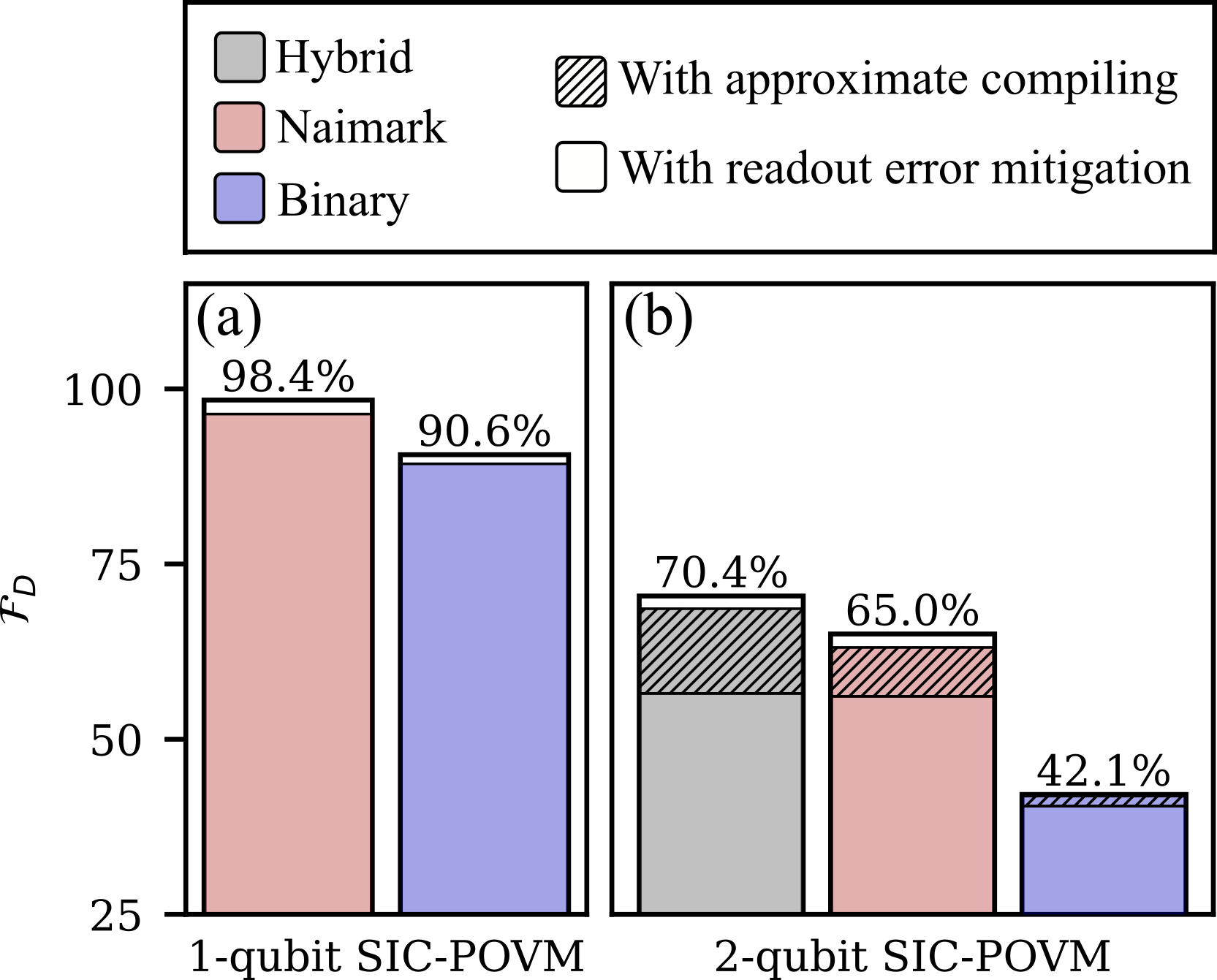}
    \caption{Comparison of the highest achieved POVM fidelities for different implementations of one- (left) and two-qubit (right) SIC-POVMs. The colored bars represent three different methods: Naimark (red), binary (blue), and hybrid (grey). White bars represent the improvement from readout error mitigation and hatched bars represent the improvement from approximate compiling. Error bars are obtained from bootstrapping and are too narrow to be visible.}
    \label{fig:detector_tomography_bar_plot}
\end{figure}

The one-qubit SIC-POVM is implemented using Naimark’s dilation and binary search, both requiring a single auxiliary qubit (see Appendix~\ref{sec:onequbitschematic}). In either case, the circuit can be exactly compiled with a small number of CNOTs. Naimark’s dilation requires 3 CNOTs and binary search results in an average CNOT depth of 4.5. Fig.~\ref{fig:detector_tomography_bar_plot} (a) summarizes the highest fidelities achieved with each method and the impact of readout error mitigation on \texttt{ibmq\_kolkata}~\cite{IBMQuantumPlatform}.
While both Naimark's dilation and binary tree require one auxiliary qubit for a single qubit SIC-POVM, there are additional sources of noise that limit the fidelity of the latter. Specifically, the binary tree applies two unitaries, which leads to a higher circuit depth. Moreover, it suffers from additional idle time due to mid-circuit readout and the delay in processing measurement outcomes to determine the next unitary operation~\cite{Baumer2023LongRange}. This reduces the fidelity of the binary tree approach. After readout error mitigation, Naimark achieves a fidelity of $98.4\pm0.2\%$ while binary tree reaches $90.6\pm0.2\%$ using CREM. Naimark's fidelity increases by 2\% due to readout error mitigation, while binary's increases by 1.3\%. The smaller enhancement from readout error mitigation in binary tree is due to two main factors. Firstly, in our experiment, the auxiliary qubit has a lower readout error than the system qubit, and therefore, the binary tree approach is less affected by those errors compared to Naimark. Secondly, qubit readout deviates from being a perfect Quantum Non-Demolition (QND) measurement~\cite{LukeGovia2022MCM}. Thus, the post-measurement state may differ from the recorded measurement outcome. Such discrepancies are not corrected by CREM as detailed in Appendix~\ref{sec:crem}.

\subsection{Two-qubit SIC-POVM}\label{sec:twoqubitsicpovm}
For the two-qubit SIC-POVM, exact compilation of required unitary operations into native gates results in circuits with a large CNOT count. This limits their applicability on near-term quantum hardware. To overcome this challenge, we use approximate compiling, which aims to find shorter circuits for a given unitary at the cost of only approximating the target unitary. In this approach, we constrain both the number and connectivity of CNOT gates within the circuit, which are interleaved with single-qubit rotation gates. An optimizer then tries to find the rotation angles for these single-qubit gates such that the resulting circuit implements the target unitary as accurately as possible \cite{Madden2022}. We use approximate compiling for all unitaries involved in the construction of each of the three schemes. For the Naimark's dilation, the optimization involves compiling a single four-qubit unitary, while for the binary search and hybrid methods, multiple three-qubit unitaries at different levels are compiled separately and then combined together to realize the measurement sequence.

\begin{figure}[!tb]
    \centering
    \includegraphics{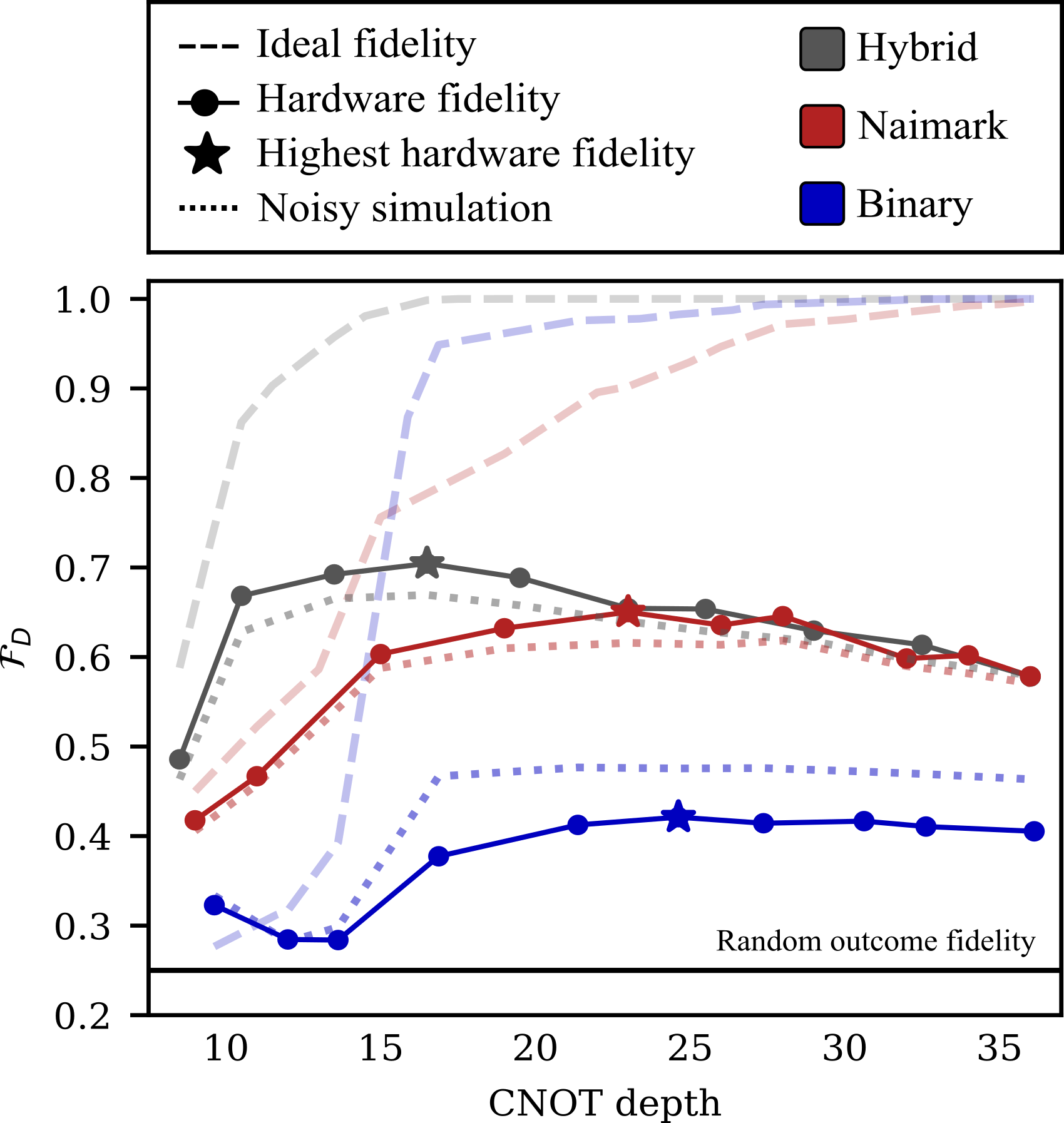}
    \caption{POVM implementation fidelity across varying CNOT depths for binary tree (blue), Naimark's dilation (red), and hybrid (grey) schemes. The line styles represent ideal fidelity (dashed), depolarizing noise model simulation (dotted), and hardware fidelity (solid), with the peak hardware fidelity denoted by a star. The fidelity of a random outcome POVM is denoted by a horizontal line at $\mathcal{F}_D$ = 0.25. Error bars are obtained from bootstrapping and are smaller than markers.}
    \label{fig:detector_tomography_combined_figure}
\end{figure}

When using approximate compiling, we expect a trade-off between circuit depth and the accuracy of the approximated POVM. For example, a low number of CNOT results in a large approximation error, as illustrated in the ideal simulation in Fig.~\ref{fig:detector_tomography_combined_figure}. However, deeper circuits will encounter more noise, and \textit{a priori} it is unclear which CNOT depth will result in the optimal performance. To approach this, we compile and run each algorithm at 10 different CNOT depths, ranging from about 9 to 35, and for each algorithm, we pick the circuit with the highest POVM fidelity in the experiment, as shown in Fig.~\ref{fig:detector_tomography_combined_figure}. Additional details of data acquisition can be found in Appendix~\ref{sec:data_acquisition}. Finally, Fig.~\ref{fig:detector_tomography_bar_plot} (b) summarizes the highest fidelities achieved on hardware with each method, along with the effect of readout error mitigation. In addition, the improvement from approximate compiling is represented by the difference between the optimal CNOT depth and the highest CNOT depth. For hybrid and Naimark, the improvement from approximate compiling is substantial. However, its effect is limited for binary tree due to long idle times in feed-forward operations, which are the main limiting factors for fidelity. We observe that the hybrid method results in the highest fidelity of $70.4\pm0.1\%$. Binary tree achieves a maximum of $42.1\pm0.1\%$ fidelity, significantly lower than the best fidelity of $65.0\pm0.1\%$ using Naimark’s dilation. However, it is still higher than the baseline fidelity of $25\%$. Even though the binary tree is more efficient in terms of required CNOT depth, its fidelity is strongly affected by the high number of feed-forward operations in the three mid-circuit measurements, causing longer idle times and coherence loss. We confirm this interpretation with numerical simulations of these experiments with a noise model inspired by the hardware in Section~\ref{sec:noise_analysis}. Finally, as in the one-qubit experiment, the lower improvement of only 0.2\% from CREM in the binary tree as compared to the 1.9\% for Naimark and 1.8\% for the hybrid circuit is attributed to the comparatively low readout errors on the auxiliary qubit as well as the non-QND errors in the auxiliary qubit readout.

\subsection{Noise analysis}\label{sec:noise_analysis}
CNOT gates are significantly noisier than single-qubit gates. Therefore, their count is a key noise metric. However, in the context of dynamic circuits, fidelity may further degrade due to idle times from mid-circuit measurements and feed-forward operations. Each conditional unitary in binary and hybrid circuits involves verifying the unitary's condition, with a time cost comparable to that of a CNOT gate's duration. 

To provide a phenomenological explanation for the two-qubit SIC-POVM results, we perform a noisy simulation using a depolarizing noise model. Firstly, to model the fidelity decay due to CNOT depth, we attach a depolarizing error $\epsilon_{\mathrm{CNOT}}=1.5\%$ to every CNOT gate. Secondly, we associate a depolarizing error $\epsilon_{\mathrm{idle}}=5\%$ with each measurement and feed-forward operation. This is motivated by the comparatively long measurement and feed-forward times, usually multiple times longer than the CNOT gate time~\cite{Baumer2023LongRange}. We choose the depolarizing errors heuristically, and the error values are consistent with recent experiments on analogous devices~\cite{chen2023Nishimori, LukeGovia2022MCM, Baumer2023LongRange}. The binary tree is impacted the most by this second kind of error because it has three mid-circuit measurements and 14 feed-forward cases, as shown in Fig.~\ref{fig:two_qubit_schematic} (d). In contrast, the hybrid only has one mid-circuit measurement and two feed-forward cases. On the other hand, Naimark's circuit requires the highest CNOT depth for approximate compilation and is, therefore, additionally affected by the approximation error, as shown in Fig~\ref{fig:detector_tomography_combined_figure}. Overall, the hybrid scheme outperforms its constituents across all CNOT depths, consistent with the experimental results. Despite the phenomenological nature of our noise model, we achieved notable agreement with the experimental data, suggesting that the error model captures the two major noise sources. An extended analysis of the performance of the three schemes in different noise regimes can be found in Appendix~\ref{sec:detailed_noise_analysis}.


\subsection{Scaling to larger systems}\label{sec:scaling_to_larger_systems}
We now discuss the resource costs of the three algorithms when scaling to larger systems.  As an illustrative example, we focus on POVMs with $M = d^2$ elements, such as informationally complete POVMs~\cite{Scott2006}, and extend this discussion in Appendix~\ref{sec:resource_estimate}. Naimark’s dilation realizes a POVM with $M = d^2$ elements through a single unitary acting on $\log_2{M} = 2n$ qubits followed by a layer of end-circuit measurements. For an $n$-qubit system, Naimark thus requires $n$ auxiliary qubits. Notably, end-circuit measurements can usually be executed in parallel, effectively counting as a single measurement step. Binary search utilizes $\log_2{M} = 2n$ layers of $(n+1)$-qubit unitaries, interleaved by $2n-1$ mid-circuit measurements and a final end-circuit measurement. Of these $2n$ unitaries, $2n-1$ are conditional. In contrast, the hybrid uses $\log_2(\frac{M}{d}) = n$ layers of $(n+1)$-qubit unitaries, interleaved by $n-1$ mid-circuit measurements and terminating with $n+1$ end-circuit measurements. Thus, the hybrid approach requires half as many layers compared to binary. Additionally, its final $n+1$ end-circuit measurements can also be parallelized, providing yet another advantage over binary. For both binary and hybrid, each mid-circuit measurement is followed by resetting the auxiliary qubit. This reset can be achieved at minimal cost by applying an X gate conditional on the measurement outcome (active reset).

We consider all unitaries in the binary search and Naimark’s dilation as generic. Therefore, an upper bound of $\mathcal{O}(4^n)$ CNOTs required for decomposing a generic $n$-qubit unitary~\cite{Iten2016Isometries} serves as a common cost unit for the circuit depth of all three schemes. Consequently, the CNOT depths for Naimark, binary, and hybrid schemes scale as \mbox{$\mathcal{O}(16^n)$}, \mbox{$\mathcal{O}(2n\cdot4^{n+1})$}, and \mbox{$\mathcal{O}(n\cdot4^{n+1})$}, respectively. We conclude that the hybrid scheme requires asymptotically the shortest circuit. Furthermore, since the number of mid-circuit measurements and conditional operations increases only linearly with system size, the CNOT depth emerges as the critical cost factor for larger systems. Finally, in Appendix~\ref{sec:detailed_noise_analysis}, we provide further analysis on the impact of noise from mid-circuit measurements and conditional operations in the two-qubit experiment.

\section{Conclusion} \label{sec:discussionconclusion}
In conclusion, we introduce a new approach for implementing single-setting POVMs on multi-qubit superconducting systems using dynamic circuits. Our method results in shorter-depth circuits and outperforms both Naimark's dilation and binary tree in implementing a two-qubit SIC-POVM. We further demonstrate that approximate compiling is an effective approach to realizing generalized measurements under noisy conditions. In addition, we devise a new CREM technique to combat error propagation in dynamic circuits and enhance the fidelity. We limit our implementation to two-qubit POVMs, as our attempts to extend this approach to three-qubit POVMs resulted in prohibitively large circuit depths in the order of hundreds of CNOTs, surpassing any previously successful implementation on similar hardware. We, therefore, expect that implementing higher-dimensional POVMs using our approach will require improvements in hardware, such as faster feed-forward and more efficient approximate compiling techniques. For example, an optimized compiling strategy could exploit the unitary freedom in the definition of Kraus operators in Eq.~\eqref{eq:kraus_operator_hybrid} or try permutations of POVM elements. Nevertheless, our results open new possibilities in the near future. For example, parallel execution of products of two-qubit POVMs would allow to cover multiple qubits on the chip, a strategy with potential implications for multi-qubit state tomography or classical shadows~\cite{Stricker2022}. We also suggest incorporating our hybrid approach in the development of hardware-efficient, parametric POVMs, which has proven to be an effective technique for targeted applications~\cite{Perez2021LearningtoMeasure}.

\textit{Note added --} During the completion of this manuscript, we became aware of a related but independently developed error-mitigation technique for mid-circuit measurements~\cite{hashim2023quasi}.

\begin{acknowledgements}
We thank the organizers of the Qiskit Advocate Mentorship Program (QAMP) by IBM Quantum, where this work was initiated. 
We also thank valuable discussions with Elisa B\"aumer, Zlatko K. Minev, Ali Javadi-Abhari, Kristan Temme, Francesco Tacchino, Thomas Alexander, Michael Healy, Brian Donovan, Luke C. G. Govia, and Moein Malekakhlagh. The views expressed are those of the authors and do not reflect the official policy or position of IBM or the IBM Quantum team. 

L.J. acknowledges support from the ARO(W911NF-23-1-0077), ARO MURI (W911NF-21-1-0325), AFOSR MURI (FA9550-19-1-0399, FA9550-21-1-0209, FA9550-23-1-0338), NSF (OMA-1936118, ERC-1941583, OMA-2137642, OSI-2326767, CCF-2312755), NTT Research, Packard Foundation (2020-71479). L.J. and A.S. thank the Kavli Institute for
Theoretical Physics (KITP) at the University of California, Santa
Barbara supported by the National Science Foundation under Grant No. PHY-1748958.

D.S.W. and A.S. contributed equally to this work.
\end{acknowledgements}


\appendix
\section{One-qubit POVM implementation}\label{sec:onequbitschematic}
\begin{figure}[!b]
    \centering
    \includegraphics{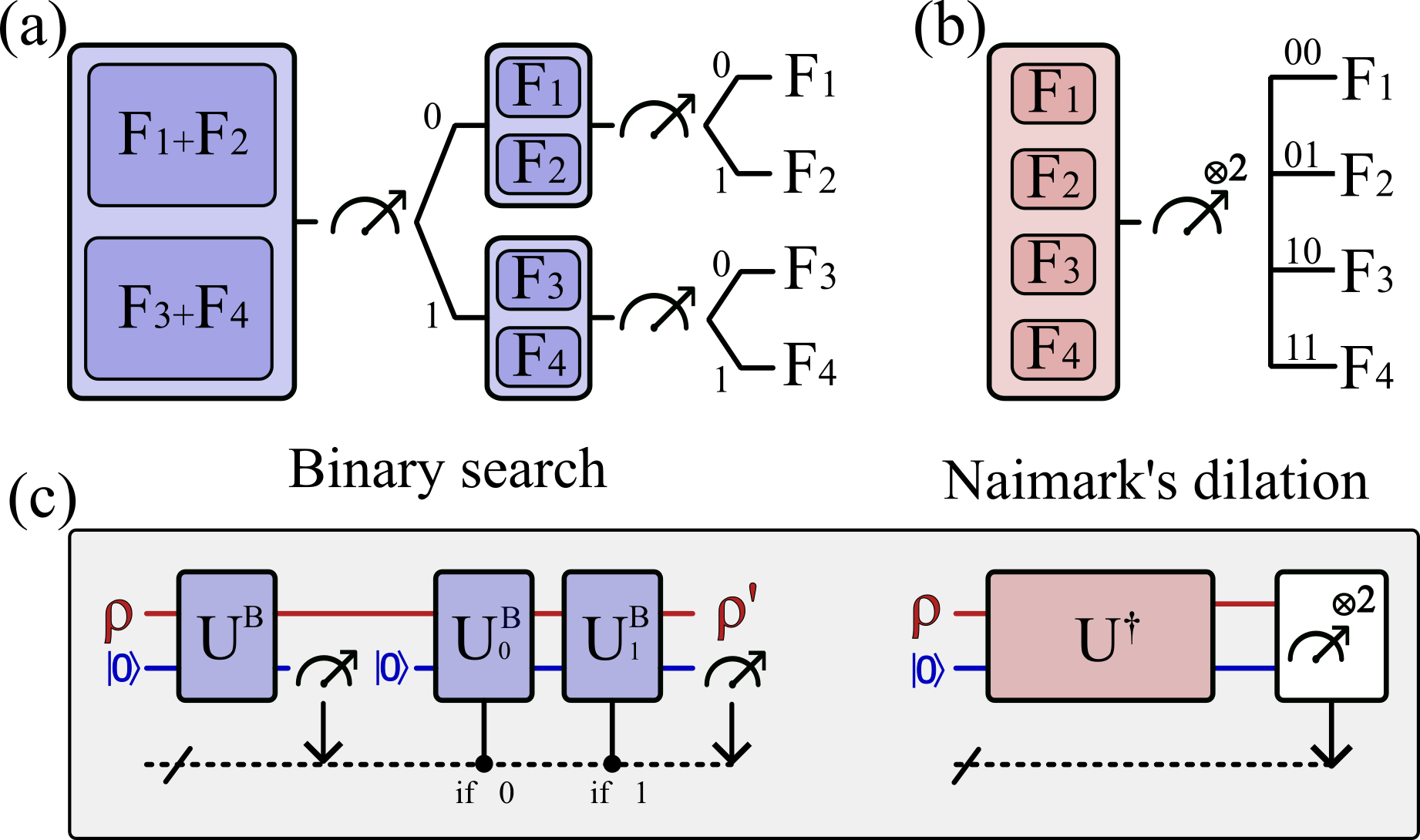}
    \caption{Schematic representation of a  single qubit POVM with $M$=4 POVM elements realized through \textbf{(a)} binary tree tree (blue), and \textbf{(b)} Naimark's dilation (red). Branching occurs after the measurement conditional on the outcome, with the trajectory defined by a bitstring of previous outcomes. Each scheme terminates with a bitstring of length two corresponding to one of 4 POVM elements. Grey panel \textbf{(c)} shows quantum circuit implementations for methods \textbf{(a)} and \textbf{(b)}. System qubits are in red, and auxiliary qubits are in blue. Binary tree includes a single mid-circuit measurement and two unitaries conditional on the classical register (dotted line).}
    \label{fig:one_qubit_schematic}
\end{figure}

A single-qubit rank-one POVM can have at most $M=4$ linearly independent elements. Using Naimark's dilation, we can realize such a POVM with a single auxiliary qubit by applying a suitable coupling unitary and measuring the system and auxiliary qubit in the computational basis, as detailed in Appendix~\ref{sec:details_naimark_dilation}. Therefore, a single-qubit POVM is a special case where the dimension of unitaries for the binary tree and Naimark's dilation coincide. In this case, we do not expect binary to offer any advantage over Naimark because it requires two layers of two-qubit unitaries, as shown in Fig.~\ref{fig:one_qubit_schematic}. Finally, our hybrid approach simplifies to a single application of Naimark's dilation because there is no need for partial filtering. 

The advantage of Naimark over binary is confirmed by the higher fidelities obtained in the one-qubit realization of a SIC-POVM in Section~\ref{sec:onequbitsicpovm}.

\section{Detector tomography} \label{sec:detector_tomography_appendix}
Detector tomography is used to reconstruct an unknown detector, or a POVM in the present work. In many aspects, it is similar to state tomography, where an unknown state is reconstructed from the measurement statistics of a known detector. In contrast, in detector tomography, we aim to reconstruct an unknown detector by preparing and measuring a set of known initial states. To elaborate, for a specific known state \(\rho_i\), the probability of getting a measurement outcome \(F_m\) is given by \(p_{mi} = \text{tr}(F_m\rho_i)\). Every $F_m$ can be expressed using multi-qubit Pauli matrices: $F_m = \sum_{k=1}^{d^2}f_{mk}\sigma_k$. This leads to the probability equation:
\begin{equation}
    p_{mi} = \sum_{k=1}^{d^2}f_{mk}\text{tr}(\sigma_k\rho_i),
\end{equation}
which is linear and analogous to state tomography. We define the matrix $S$ as the matrix containing trace values that are specific to the set of initial states:
\begin{equation}
S = \begin{pmatrix}
    \text{tr}(\sigma_1\rho_1) & \cdots & \text{tr}(\sigma_{d^2}\rho_1) \\
    \text{tr}(\sigma_1\rho_2) & \cdots & \text{tr}(\sigma_{d^2}\rho_2) \\
    \vdots & \vdots & \vdots \\
    \text{tr}(\sigma_1\rho_n) & \cdots & \text{tr}(\sigma_{d^2}\rho_n)
\end{pmatrix}.
\end{equation}
Then, by writing $\vec{f}_m$ as the vector of coefficients in front of Pauli matrices in the expansion of $F_m$ and $\vec{p}_m$ as the vector of corresponding probabilities determined from the experiment, we find the solution to $\vec{p}_m = S\vec{f}_m$ by computing the least-squares estimate:
\begin{equation}
\vec{f}_m = (S^{\dagger}S)^{-1}S^{\dagger}\vec{p}_m.
\end{equation}

The procedure above has to be repeated for every unknown POVM element $F_m$. In general, the least-squares estimate is not guaranteed to produce a non-negative $F_m$, especially with few measurement samples. Therefore, the Choi matrix in Eq.~\eqref{eq:normalized_choi_matrix} of the reconstructed POVM may have small negative eigenvalues. To impose the non-negativity, one can rescale the eigenvalues of the Choi matrix from least-squares, effectively projecting it onto the set of quantum states~\cite{Smolin2012QST,Guta2020PLS}. This procedure corresponds to computing the maximum likelihood estimate under the assumption of Gaussian noise~\cite{Smolin2012QST}. In our experiments, however, we take sufficient measurement samples such that rescaling is only necessary for a few POVM elements. In particular, rescaling has no effect on fidelities within the statistical uncertainty.

For a two-qubit experiment, 36 Pauli basis states were used, making the set of initial states overcomplete. Consequently, the matrix $S$ had dimensions $36\times16$. We also used an overcomplete set of 6 Pauli basis states for a one-qubit experiment, resulting in a $6\times4$ matrix. In principle, only $d^2$ initial states are fundamentally required to form a complete basis. However, using extra states improves the estimation accuracy and compensates for the state-dependent performance of a particular POVM implementation.

\section{Conditional readout error mitigation} \label{sec:crem}
\begin{figure}[!tb]
    \centering
    	\includegraphics{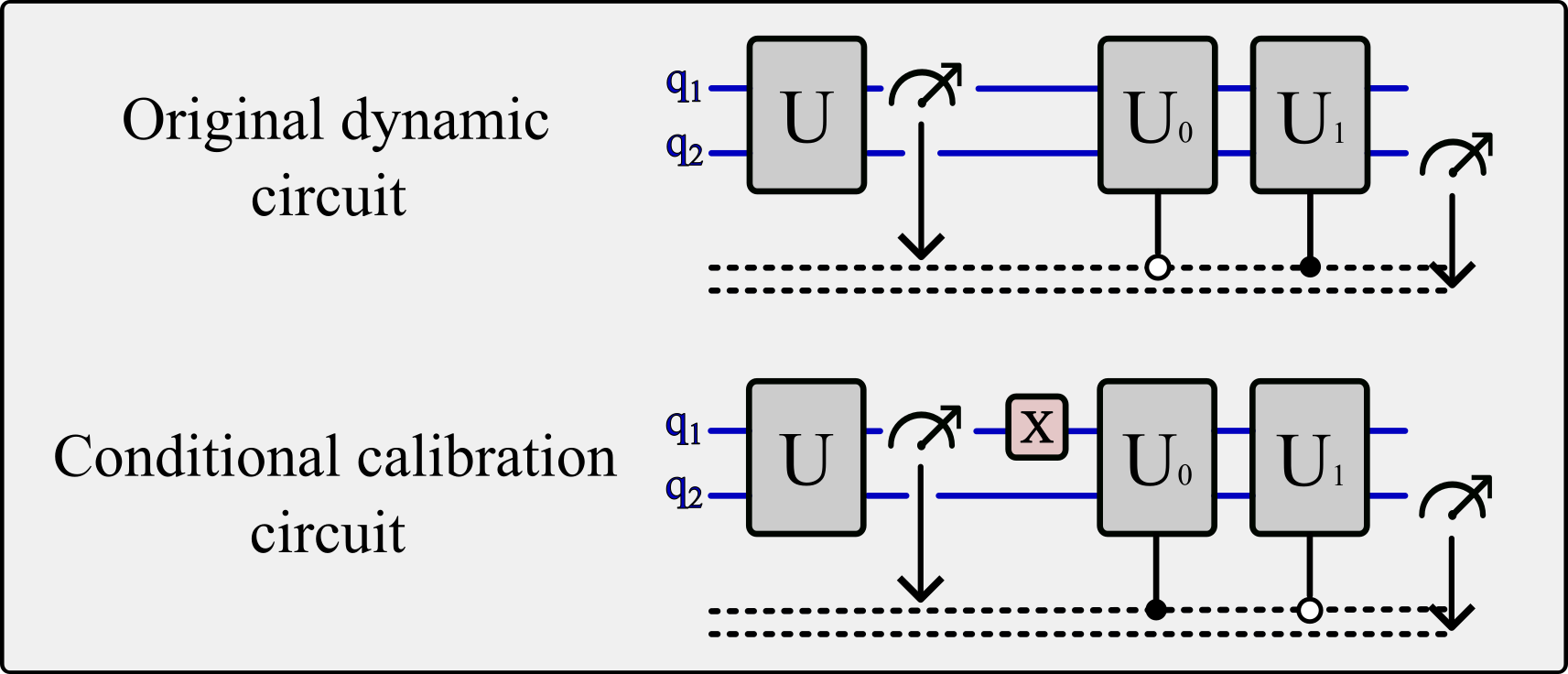}
    \caption{Conditinal readout error mitigation: (a) Original two-qubit circuit with a unitary applied to $q_1$ and $q_2$, followed by a mid-circuit measurement, with subsequent operations conditioned on the measurement's outcome. (b) Conditional calibration circuit replicating the original with the feed-forward conditions exchanged, and the post-measurement state flipped.}
    \label{fig:crem_schematic}
\end{figure}

Imprecise multi-qubit measurements can be described using a classical probabilistic model. Measurement errors arise from random misclassification of correct outcomes, represented by a confusion matrix $M$, where $M_{ij}$ = $\mathrm{p(i|j)}$ is the probability of misclassifying $j$ as $i$. Here, $i$ and $j$ are computational basis states. Observed probabilities $(P)$ are obtained by applying the confusion matrix to ideal probabilities $(Q)$: $P = MQ$. For example, for the case of measuring one qubit $(i\in {0, 1})$, the confusion matrix is given as:
\begin{equation}
    M = 
        \left(
        \begin{array}{c c}
            1-\epsilon_0 & \epsilon_1 \\
            \epsilon_0 & 1-\epsilon_1 \\
        \end{array}
        \right).
    \label{eq:one_qubit_confusion_matrix}
\end{equation}
For instance, the probability of obtaining $0$ is the sum of the probability $(1-\epsilon_0)$ of correctly identifying $0$ as $0$ and the probability $\epsilon_1$ of misclassifying $1$ as $0$. Error-free probabilities $Q$ can be obtained by inverting the confusion matrix: $Q = M^{-1}P$.
In general, measuring $n$ qubits in the computational basis can result in any of $2^n$ computational states with some unique probability. Therefore, to characterize the full confusion matrix $M$, one needs to obtain $4^n$ matrix elements through calibration. Often, however, it is justified to assume that readout errors are local, i.e., the probability of misclassifying the state of some qubit is independent of the states of other qubits. This allows us to construct the global confusion matrix $M$ as the tensor product of $n$ $2 \times 2$
confusion matrices $M_i$, one for each qubit:
\begin{equation}
M = M_n \otimes M_{n-1} \otimes \ldots \otimes M_1.
\end{equation}
The situation is different with dynamic circuits, which enable mid-circuit measurements, allowing subsequent gates to be conditional on these measurement outcomes.
For instance, consider the two-qubit circuit in the Fig.~\ref{fig:crem_schematic}. A unitary operation $U$ prior to the first mid-circuit measurement encompasses the circuit's unitary segment. After measuring qubit $q_1$, the result is stored in the first bit of the classical register. Subsequently, the circuit applies $U_0$ if the measurement reports $0$ and $U_1$ if $1$. We see that if a bit-flip error occurs during the mid-circuit measurement, then not only is the wrong measurement outcome recorded, but also the wrong condition is triggered, and the wrong unitary is applied. Moreover, the unitaries now act on the flipped post-measurement state of qubit $q_1$. In this way, the measurement error from mid-circuit measurement propagates through the circuit, affecting the evolution of the quantum state.
To account for this error propagation, consider the second circuit in Fig.~\ref{fig:crem_schematic}. The second circuit, which we call a conditional calibration circuit, differs from the original circuits in that the conditions of the two unitaries are inverted, and an $X$ gate flips the post-measurement state. Notice that if a bit-flip occurs during the mid-circuit measurement in the second circuit, the subsequent state evolution will correspond to the case when no bit-flip occurs in the first circuit. Denote $\epsilon_i^k$ as the probability of misclassifying the state $i$ on qubit $q_k$; $P$ and $Q$ as the noisy and ideal probabilities for the original circuit; and $\tilde P$ and $\tilde Q$ as the corresponding probabilities for the conditional calibration circuit. Also, $P_{ij}$ denotes the probability of measuring $q_1$ in state $j$ and $q_2$ in state $i$. We can now express the noisy probability $P_{00}$ of obtaining the outcome $00$ as:
\begin{align*}
    P_{00} &= (1-\epsilon_0^2)(1-\epsilon_0^1) Q_{00} + (1-\epsilon_0^2)\epsilon_1^1 \tilde Q_{01} \\
    & + \epsilon_1^2(1-\epsilon_0^1) Q_{10} + \epsilon_1^2\epsilon_1^1 \tilde Q_{11}.
\end{align*}
We see that the noisy probability $P_{00}$ includes four different scenarios:
\begin{enumerate}
    \item The ideal case scenario when no readout error occurs during both mid-circuit and end-circuit measurement with probability $(1-\epsilon_0^2)(1-\epsilon_0^1)$.
    \item The outcome $01$ is misclassified as $00$ with probability $(1-\epsilon_0^2)\epsilon_1^1$ due to the error on the mid-circuit measurement. However, the subsequent state evolution corresponds to the error-free scenario. 
    \item The outcome $10$ is misclassified as $00$ with probability $\epsilon_1^2(1-\epsilon_0^1)$ due to the error on the end-circuit measurement.
    \item The outcome $11$ is misclassified as $00$ with probability $\epsilon_1^2\epsilon_1^1$ due to errors on both measurements, whereby the effective state evolution is unaffected by measurement errors.
\end{enumerate}
In a similar manner we can obtain expressions for $P_{01}$, $P_{10}$ and $P_{11}$:
\begin{align*}
    P_{01} &= \epsilon_0^1 (1 - \epsilon_0^2) \tilde Q_{00} + (1 - \epsilon_0^2) (1 - \epsilon_1^1) Q_{01} \\
    &+ \epsilon_0^1 \epsilon_1^2 \tilde Q_{10} + (1 - \epsilon_1^1) \epsilon_1^2 Q_{11} \\
    P_{10} &= (1 - \epsilon_0^1) \epsilon_0^2 Q_{00} + \epsilon_0^2 \epsilon_1^1 \tilde Q_{01} \\
    &+ (1 - \epsilon_0^1) (1 - \epsilon_1^2) Q_{10} + \epsilon_1^1 (1 - \epsilon_1^2) \tilde Q_{11} \\
    P_{11} &= \epsilon_0^1 \epsilon_0^2 \tilde Q_{00} + \epsilon_0^2 (1 - \epsilon_1^1) Q_{01} \\
    &+ \epsilon_0^1 (1 - \epsilon_1^2) \tilde Q_{10} + (1 - \epsilon_1^1) (1 - \epsilon_1^2) Q_{11}.	
\end{align*}
Following the same logic, we obtain the expression for $\tilde P_{00}$, $\tilde P_{01}$, $\tilde P_{10}$ and $\tilde P_{11}$. It is convenient to write:
\begin{align*}
    \begin{pmatrix}
        P_{00} \\
        \tilde P_{01} \\
        P_{10} \\
        \tilde P_{11}
    \end{pmatrix}
    &=
    \begin{bmatrix}
    \begin{pmatrix}
        1 - \epsilon_0^1 & \epsilon_1^1 \\
        \epsilon_0^1 & 1 - \epsilon_1^1
    \end{pmatrix}
    \bigotimes
    \begin{pmatrix}
        1 - \epsilon_0^2 & \epsilon_1^2 \\
        \epsilon_0^2 & 1 - \epsilon_1^2
    \end{pmatrix}
    \end{bmatrix}
    \begin{pmatrix}
        Q_{00} \\
        \tilde Q_{01} \\
        Q_{10} \\
        \tilde Q_{11}
    \end{pmatrix} \\
    &=
    \begin{bmatrix}
        M_2
        \bigotimes
        M_1
    \end{bmatrix}
    \begin{pmatrix}
        Q_{00} \\
        \tilde Q_{01} \\
        Q_{10} \\
        \tilde Q_{11}
    \end{pmatrix}
\end{align*}
and
\begin{align*}
    \begin{pmatrix}
        \tilde P_{00} \\
        P_{01} \\
        \tilde P_{10} \\
        P_{11}
    \end{pmatrix}
    &=
    \begin{bmatrix}
        \begin{pmatrix}
            1 - \epsilon_0^1 & \epsilon_1^1 \\
            \epsilon_0^1 & 1 - \epsilon_1^1
        \end{pmatrix}
        \bigotimes
        \begin{pmatrix}
            1 - \epsilon_0^2 & \epsilon_1^2 \\
            \epsilon_0^2 & 1 - \epsilon_1^2
        \end{pmatrix}
    \end{bmatrix}
    \begin{pmatrix}
        \tilde Q_{00} \\
        Q_{01} \\
        \tilde Q_{10} \\
        Q_{11}
    \end{pmatrix} \\
    &=
    \begin{bmatrix}
        M_2
        \bigotimes
        M_1
    \end{bmatrix}
    \begin{pmatrix}
        \tilde Q_{00} \\
        Q_{01} \\
        \tilde Q_{10} \\
        Q_{11}
    \end{pmatrix}.
\end{align*}
Here, $M_i$ denotes the confusion matrix on qubit $q_i$. In a similar manner to the standard readout error mitigation, we can invert the Kronecker product $(M_2 \bigotimes M_1)^{-1}$ = $M_2^{-1} \bigotimes M_1^{-1}$ to obtain error-free probabilities $Q$. As a by-product, we also obtain error-free probabilities $\tilde Q$ for the conditional calibration circuit. It is worth noting that the described post-processing requires obtaining probabilities for two circuits (original and calibration circuit), which doubles the total number of samples.

\begin{figure}[!t]
    \centering
    \includegraphics{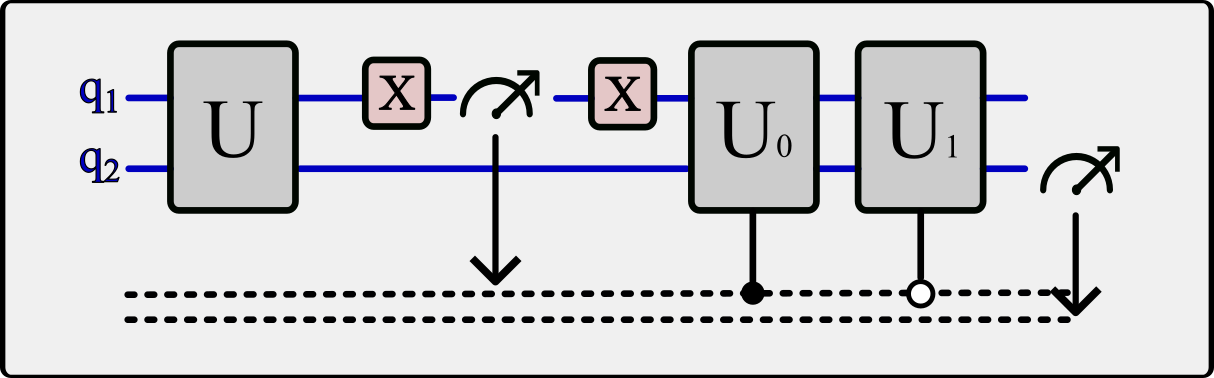}
    \caption{Two-qubit circuit modeling non-QND readout errors as post-measurement bit-flip with probability $\epsilon_{\mathrm{QND}}$. The ``correctable" error is modeled as a pre-measurement bit-flip with probability $\epsilon$.}
    \label{fig:crem_random_circuit}
\end{figure}

\begin{figure}[!hb]
    \centering
    \includegraphics{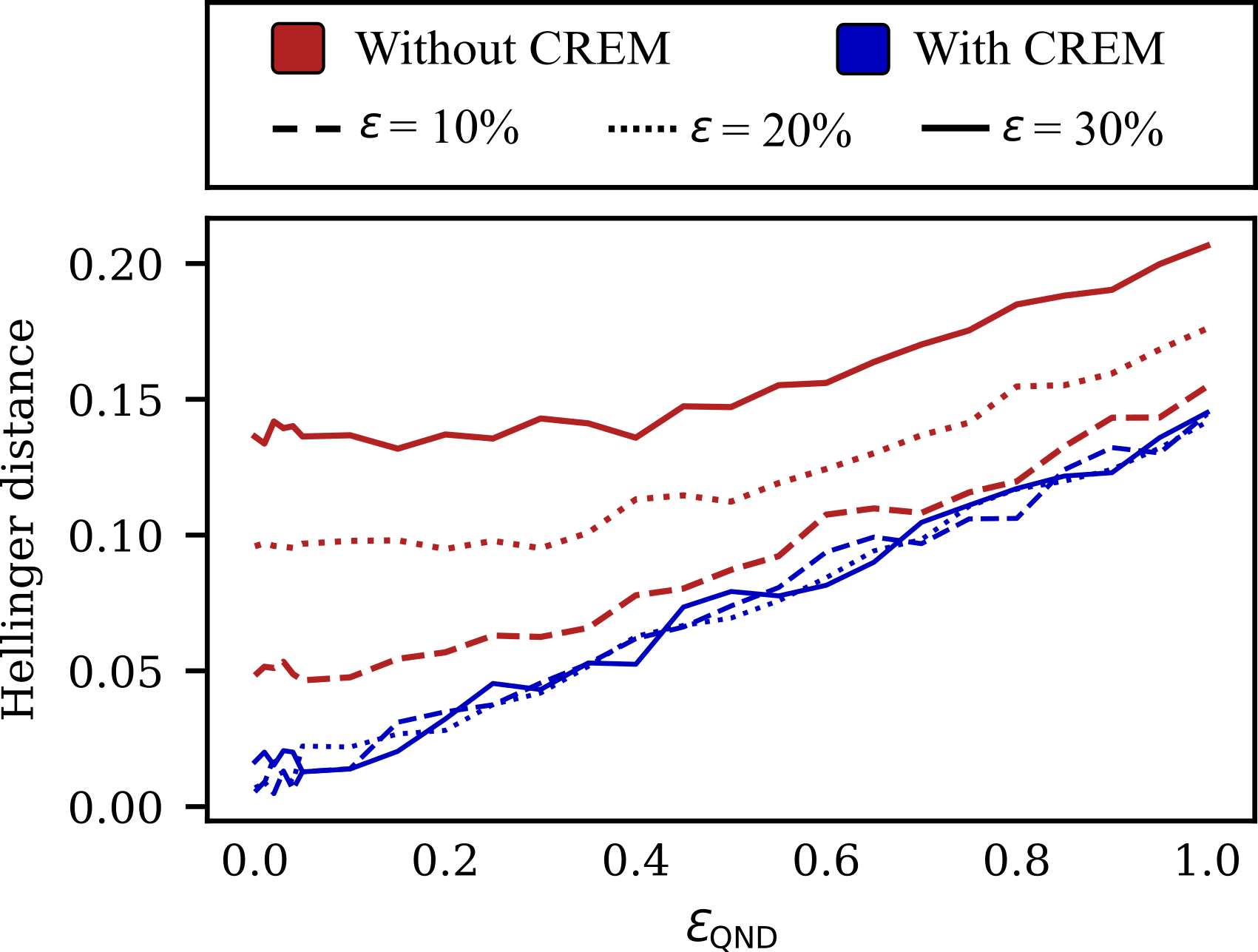}
    \caption{Illustration of the increase in Hellinger distance between the noisy and noiseless probability distributions with the rise of non-QND error rate $\epsilon_{\mathrm{QND}}$, comparing outcomes with (blue) and without (red) CREM across different ``correctable" error rates $\epsilon$. In the presence of both errors, CREM removes the contribution of ``correctable" errors.}
    \label{fig:crem_analysis}
\end{figure}

The described procedure extends to multiple mid-circuit measurements. The number of required conditional calibration circuits (including the original circuit) is given by the number of possible combinations of measurement errors, which is  $2^{n_\mathrm{mid}}$ where $n_\mathrm{mid}$ is the number of mid-circuit measurements. Therefore, the sampling overhead scales exponentially with $n_\mathrm{mid}$, which makes this technique feasible only for a few numbers of mid-circuit measurements. Note that the underlying assumption we made about the nature of readout errors is that readout is perfectly QND. In this setting, CREM can correct readout errors. In reality, the readout of qubits is not perfectly QND, which can result in inconsistencies between the post-measurement state and the measurement outcome. For example, a qubit can decay during the measurement pulse. Moreover, applying measurements can cause leakage from the computation subspace to other states outside of the computational space. These errors contribute to non-QND measurement errors, which alter the state after the measurement~\cite{LukeGovia2022MCM}. In mid-circuit measurements, these errors affect the subsequent computations, whereas in final measurements they manifest as initialization errors in the next round. CREM does not account for non-QND errors. To illustrate how CREM performs in the presence of non-QND errors, we use a simple noise model that includes random bit-flips before and after the noiseless measurement pulse. The corresponding two-qubit circuit is shown in Fig~\ref{fig:crem_random_circuit}. The random bit-flip before the measurement occurs with probability $\epsilon$ and represents the error that misclassifies the qubit state but leaves the post-measurement state consistent with the measurement outcome. The random bit-flip with probability $\epsilon_{\mathrm{QND}}$ after the measurement represents the non-QNDness of the readout. The measurement on the second qubit is ideal, and the unitaries in the circuit are chosen randomly. 
By sweeping the non-QND error $\epsilon_{\mathrm{QND}}$, as shown in Fig.~\ref{fig:crem_analysis}, we observe that the Hellinger distance between the noisy and ideal measurement outcome distributions increases for both mitigated and unmitigated results. However, the error of unmitigated results also grows with $\epsilon$, while mitigated results are independent of $\epsilon$. Therefore, in a realistic setting, where both kinds of errors are present, CREM takes out the contribution of the first type of error.

\section{Resource estimate}\label{sec:resource_estimate}
In Section~\ref{sec:scaling_to_larger_systems} of the main text, we discuss the upper bound on the number of CNOTs for all three schemes for POVMs with $M = d^2$. Here, we extend the discussion to POVMs with $M \le d^2$, still assuming that POVM elements are linearly independent rank-one operators. As follows from the main text, Naimark's dilation requires a single $\lceil\log_2{M}\rceil$-qubit unitary. The binary tree scheme requires $n+1$-qubit unitaries to be implemented $\lceil\log_2{M}\rceil$ times. Finally, the hybrid scheme requires $n+1$-qubit unitaries to be implemented $\lceil\log_2{\frac{M}{2^n}}\rceil$ times. Table~\ref{tab:cnot_bounds} provides asymptotic upper bounds on the number of CNOTs.
	\begin{table}[!tb]
		\centering
		\begin{tabular}{|c|c|c|c|}
			\hline
			$M$ & Naimark & Binary & Hybrid\\
			\hline
			\hline
			$ 2^n < M \le 2^{n+1}$ & $4^{n+1}$ & $(n+1)\cdot4^{n+1}$ & $4^{n+1}$\\
			\hline
			$2^{n+1} < M \le 4^n $ & $M^2$ & $\log_2{M}\cdot4^{n+1}$ & $(\log_2{M}-n)\cdot4^{n+1}$ \\
			\hline
			\multicolumn{4}{c}{} \\
			\multicolumn{4}{c}{Example for a minimal IC-POVM} \\
			\hline
			$M = 4^n$ & $16^n$ & $2n\cdot4^{n+1}$ & $n\cdot4^{n+1}$ \\
			\hline
		\end{tabular}
		\caption{Upper bounds on the number of CNOTs required for POVM implementation.}
		\label{tab:cnot_bounds}
	\end{table}
For clarity, we treat the case when $2^n < M \le 2^{n+1}$ separately. We first pad the POVM with zero operators to the nearest power of two. That is, if a given POVM initially has $2^n < M < 2^{n+1}$ elements, we pad it until $M = 2^{n+1}$. The POVM with $M = 2^{n+1}$ elements can be most efficiently implemented with Naimark's dilation using a single auxiliary qubit. Binary search requires $(n+1)$ partial filtering steps, while the hybrid scheme coincides with Naimark's dilation, as in the case of one-qubit SIC-POVM in Appendix~\ref{sec:onequbitsicpovm}. An important example is a minimal informationally complete POVM, i.e., a POVM with $4^n$ linearly-independent POVM elements, such as SIC-POVM. In this case, the hybrid scheme results in a circuit with half the length compared to the binary search, as shown in Table~\ref{tab:cnot_bounds}.
	
In summary, the hybrid scheme yields a shorter circuit than the binary search tree for any system size and number of POVM elements. It outperforms Naimark's dilation for $M > 2^{n+1}$ while coinciding with Naimark's dilation when $M \le 2^{n+1}$.

\section{State tomography} \label{sec:state_tomography}
Quantum state tomography is one of the prominent examples where the choice of measurement plays a central role. A typical approach to perform state reconstruction of a multi-qubit system is to perform tomographic measurements on each individual qubit using each of the Pauli bases. For an $n$-qubit system, this requires $3^n$ measurement settings, a number which grows exponentially with the system size, making it impractical for large numbers of qubits. Moreover, Pauli bases are known to be sub-optimal for state tomography, requiring more samples than necessary due to the informational redundancy of its projectors. The last aspect is investigated in more detail in Appendix~\ref{sec:sic_vs_pauli}.

SIC-POVMs, in contrast, are known to be sample-optimal for state tomography~\cite{Renes2004}. In the previous sections, we performed the detector tomography by assuming perfect knowledge of the prepared states to obtain the tomographic information about the implemented POVM. Conversely, we could assume the perfect knowledge of the detector (SIC-POVM) to reconstruct the prepared quantum states. Of course, since the implemented POVM is imperfect, the resulting reconstruction fidelity will deviate from the ideal. 
\begin{figure}[!tb]
    \centering
    \includegraphics{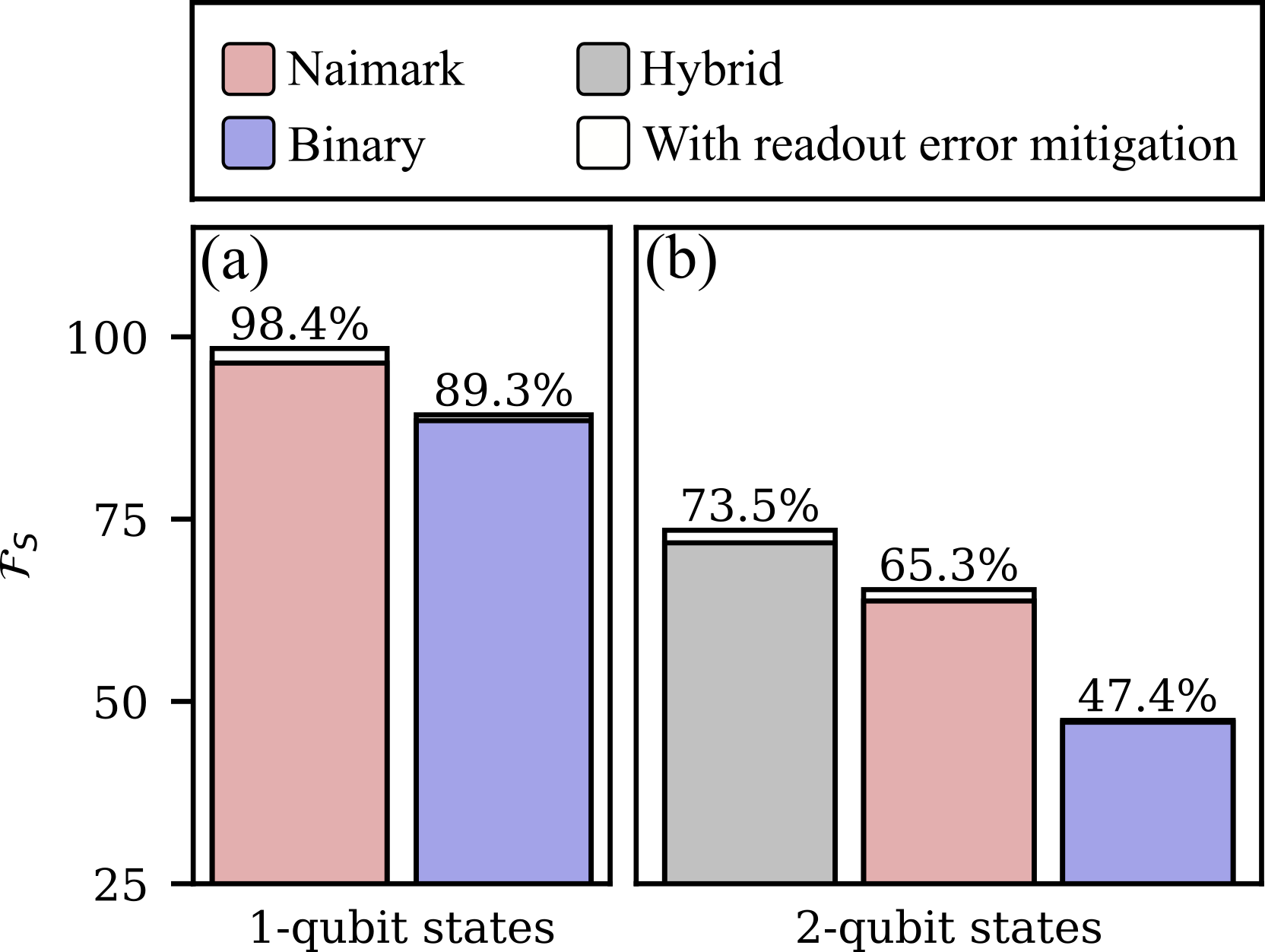}
    \caption{Comparison of the highest achieved (average) state fidelities for different implementations of one- (left) and two-qubit (right) SIC-POVMs. The colored bars represent three different methods: Naimark (red), binary (blue), and hybrid (grey). White bars represent the improvement from readout error mitigation.}
    \label{fig:state_tomography_bar_plot}
\end{figure}

Notably, we don’t need to perform any additional measurements to the ones obtained for detector tomography. We simply recast the problem and use the ideal SIC-POVM and the measurement statistics to perform the linear inversion. In practice, the reconstructed states may not be physical due to small negative eigenvalues. To address this problem without significant computational overhead, we rescale the eigenvalues of each reconstructed density matrix to make it positive semi-definite~\cite{Smolin2012QST}. Fig.~\ref{fig:state_tomography_bar_plot} shows the average reconstruction fidelities of the 6 single-qubit Pauli basis states and the 36 two-qubit Pauli basis states. Here, for the two-qubit experiment, we used the optimal CNOT depths determined in Section~\ref{sec:experiment}. We see that, overall, the achieved peak fidelities by each algorithm agree closely with the corresponding POVM fidelities.

\section{Resource overhead of state tomography} \label{sec:sic_vs_pauli}
In the context of state tomography with Pauli basis, we need $3^n$ different measurement settings, each yielding $d$ outcomes. The total number of measurement outcomes exceeds the $d^2-1$ parameters of a $d$-dimensional density matrix. Therefore, Pauli bases provide redundant information, which results in a higher variance in the fidelity of the reconstructed state~\cite{Bent2015SICPOVM}. 

\begin{figure}[!htb]
    \centering
    \includegraphics{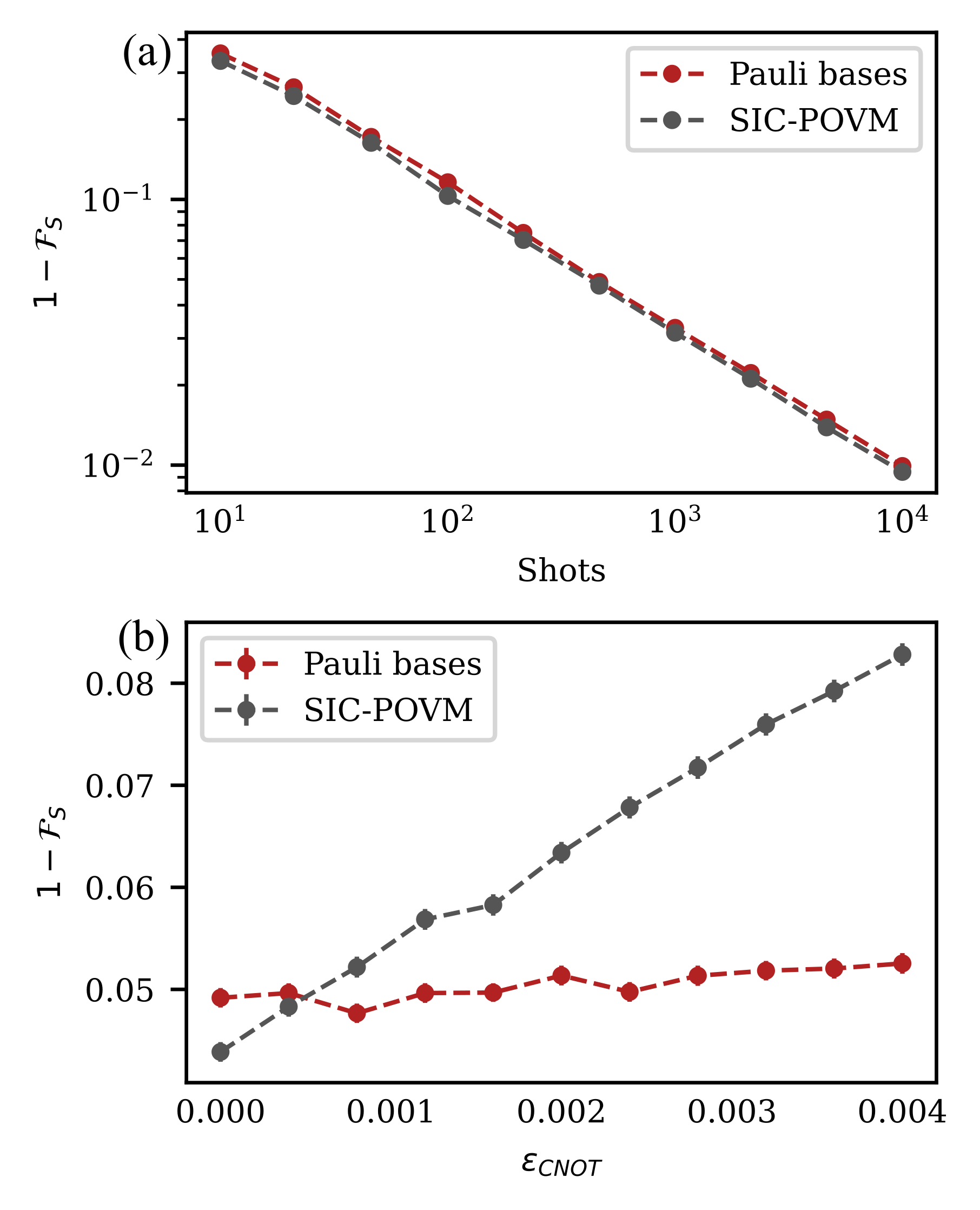}
    \caption{Comparison of the infidelity of state reconstruction between Pauli bases (red) and SIC-POVM (grey) as a function of the number of measurement shots (a) and the CNOT gate error rate (b). In (a), the simulation is noiseless, and in (b), the number of shots is fixed at 500.}
    \label{fig:bias_variance}
\end{figure}

SIC-POVMs, however, are proven optimal for state tomography~\cite{Renes2004}. This is observed in Fig.~\ref{fig:bias_variance} by comparing the infidelity of reconstruction for a two-qubit SIC-POVM with varying numbers of shots using a noiseless Qiskit simulator. At lower shots, SIC-POVM exhibits a slight advantage, around 1\% at 100 shots and 0.15\% at 1000 shots. As the number of shots increases, both methods converge to zero infidelity at a rate of $N^{\frac{1}{2}}$, evident from the -1/2 slope of the curves.
However, due to noise in the circuit that implements a SIC-POVM, its performance is considerably lower than that of the Pauli bases. Nevertheless, if the noise level of the device is sufficiently low, using SIC-POVMs for state tomography will be more efficient.

We investigate the performance of SIC-POVM under a simplistic noise model that applies depolarizing noise to the CNOT gates. The purpose of this analysis is to showcase a crossover point where SIC-POVM reconstruction yields a higher fidelity than Pauli bases for a fixed number of 500 shots. The Fig.~\ref{fig:bias_variance} shows the corresponding tradeoff. At low depolarizing noise, SIC-POVM results in lower infidelity, which grows as we increase the error. Pauli basis tomography is expectedly unaffected by this noise model because Pauli basis measurements don't include any CNOTs. A more realistic noise model would include single-qubit and measurement errors.

\section{Details of the Naimark’s dilation}\label{sec:details_naimark_dilation}
Any POVM can be realized by a projective measurement in a higher-dimensional Hilbert space by introducing a number of auxiliary qubits~\cite{Naimark1976}. In general, the dimension of the required auxiliary system dimension will depend on the number of POVM elements, the dilation method, and the ability to perform direct measurements on the system~\cite{Chen2007}. One way to extend the Hilbert space is by, for example, coupling the qubit system to neighboring qubits on a superconducting chip. Moreover, in the scope of this paper, we are only interested in the measurement statistics $\mathrm{P(i)}$ and disregard the post-measurement state of the system. Therefore, we consider the situation where both the system and auxiliary qubit are measured directly. The extension of the Hilbert space requires a minimal number of $n_A$ auxiliary qubits so that the dimension of the compound system $2^{n+n_A} \ge M$. Therefore, the auxiliary qubit resource is best utilized when $M$ is a power of two. Otherwise, we have to pad our set of POVM elements with zero operators until $M$ is a power of two. For a POVM of $M$ elements we find $M$ unnormalized vectors $\ket{\psi_i}$ such that $F_i = \ket{\psi_i}\bra{\psi_i}$. These vectors can be arranged as column vectors to form an $d\times M$ array whose rows are orthonormal $M$-dimensional vectors. This fact follows from the completeness of POVM: $\sum_{i=1}^{M}F_i = I$. Such an array can always be extended to a $M\times M$ unitary matrix, as shown in Eq.~\eqref{eq:naimark_matrix}:
\begin{equation}
    U = 
        \left(
        \begin{array}{c c c}
            \psi_{11} & .. & \psi_{M1}\\
            : &    & : \\
            \psi_{1d} & .. & \psi_{Md}\\
            * & .. & *\\
        \end{array}
        \right) \in \mathbb{U}(M).
        \label{eq:naimark_matrix}
\end{equation}
The circuit implementing Naimark's dilation is constructed by preparing the auxiliary qubit in the $\ket{0}^{\otimes n_A}$ state and applying the coupling unitary $U^{\dagger}$. Finally, by projectively measuring the compound system in the computational basis after the unitary transformation $U^{\dagger}$, we obtain the outcomes $\ket{\psi_i^{\mathrm{ext}}}\bra{\psi_i^{\mathrm{ext}}}$, where $\ket{\psi_i^{\mathrm{ext}}}$ are the columns of $U^{\dagger}$. The probability of observing an outcome $i$ is given by taking the trace over the auxiliary qubits and the system:
\begin{align*}
    \mathrm{P(i)} &= \tr(\ket{\psi_i^{\mathrm{ext}}}\bra{\psi_i^{\mathrm{ext}}}(\rho_A\otimes\rho)) \\
    &= \tr(\ket{\psi_i}\bra{\psi_i} \rho) \\
    &= \tr(F_i\rho).
\end{align*}
With this, we fully recover the original POVM measurement statistics  $\mathrm{P(i)} = \tr(F_i\rho)$.

Finally, we illustrate the procedure in Fig.~\ref{fig:one_qubit_schematic} (b) for a 4-element POVM on a single-qubit system with the corresponding circuit depicted in Fig.~\ref{fig:one_qubit_schematic} (c). Moreover, an example of Naimark's dilation applied to a two-qubit system to realize a 16-element POVM is illustrated in Fig.~\ref{fig:two_qubit_schematic} (c) with the circuit shown in Fig.~\ref{fig:two_qubit_schematic} (d).

In conclusion, it is worth noting that the way in which the dilation is realized can be different. For example, suppose the system of interest harbors more dimensions than the dimension of its computational subspace, such as higher-energy states of superconducting qubits~\cite{Fischer2022} or trapped ions~\cite{Stricker2022}. In that case, the extended Hilbert space can be a direct sum $\mathcal{H}_{\mathrm{ext}} = \mathcal{H} \oplus \mathcal{H}_A$ of the system and auxiliary spaces~\cite{Chen2007, Preskill97}. Often, such dilation is rather difficult because one needs to discriminate between qudit states efficiently.

\section{Details of the binary tree construction}\label{sec:details_binary_tree}
In this section, we outline the key steps of the binary tree protocol using the notation from Shen et al~\cite{Shen2017}. A detailed treatment can be found in Andersson et al.~\cite{Andersson2008} and Shen et al.~\cite{Shen2017}. To construct the binary search tree, we begin by padding our set of POVM elements with zero operators until $M$ is the nearest power of two. At the first level $l=1$, we find a suitable Kraus operator $A_0$ for the coupling unitary from the diagonalization of $B_0 =V_0 D_0^2 V_0^{\dagger}$, resulting in $A_0 = V_0 D_0 V_0^{\dagger}$. Here, $D_0$ is a diagonal matrix with non-negative eigenvalues because $B_0$ is positive Hermitian. We analogously construct $A_1 = V_1 D_1 V_1^{\dagger}$ from $B_1 =V_1 D_1^2 V_1^{\dagger}$. As detailed in the main text, we implement this binary POVM via an indirect measurement of the system by constructing a suitable coupling unitary.

After the initial two-outcome POVM, the post-measurement state of the system, denoted as $\Rho{1}$, will depend on the measurement outcome. For instance, if the auxiliary qubit is measured in the $\ket{0}$ state, the post-measurement state $\rho_0$ can be expressed as:
\begin{equation}
    \rho_0 = \frac{A_0\rho A_0^{\dagger}}{\tr(A_0\rho A_0^{\dagger})},
\label{eq:post_measurement_state}
\end{equation}
where $\rho$ represents the initial state of the system. Therefore, the measurement of the auxiliary qubit causes branching, effectively performing partial filtering. This procedure may be seen as a quantum instrument that combines the classical measurement outcome with the conditional post-measurement quantum state of the system~\cite{Wilde2016QuantumInfo}. The subsequent binary POVMs must take this post-measurement state into account. Therefore, we modify the Kraus operators for all subsequent steps $l\ge2$ as follows:
\begin{equation}
    \A{l} = \K{l} \K{l-1}^+ + \frac{1}{\sqrt{2}}\Q{l-1},
    \label{eq:binary_kraus_operators}
\end{equation}
where $\K{l} = \sqrt{\sum_{i=a}^{b}F_i}$ is obtained by aggregating the POVM elements located in the last level of the branch that starts from $b^{(l)}$ with indices ranging from $a$ to $b$. The choice of $\K{l}$ is not unique because an arbitrary unitary transformation $\K{l} \rightarrow \W{l}\K{l}$ leaves the measurement statistics invariant. In contrast to the construction of Kraus operators for the hybrid scheme, $\K{l-1}$ will not be invertible if the number of remaining non-zero POVM elements is smaller than the system's dimension. Therefore, $\K{l-1}^+$ denotes the Moore-Penrose pseudo-inverse of $\K{l-1}$, and $\Q{l-1}$ ensures that the pair of binary Kraus operators generates a complete POVM.

Overall, the construction of corresponding binary Kraus operators in Eq.~\eqref{eq:binary_kraus_operators} follows a sequential process:
\begin{enumerate}
    \item First, we form $\B{l} = \sum_{i=a}^{b}F_i$ by aggregating the POVM elements located in the leaves of the branch that starts from $b^{(l)}$. This results in $\B{l} = \V{l}\D{l}^2\V{l}^{\dagger}$.
    \item Next, we create $\K{l} = \V{l}\D{l}\V{l}^{\dagger}$ that satisfies the condition $\K{l}^{\dagger}\K{l} = \B{l}$.
    \item We compute $\K{l-1}^+$, which represents the Moore-Penrose pseudo-inverse of $\K{l-1}$ which was obtained in the previous level.
    \item Then, we construct the support projection matrix of $\D{l-1}$, denoted as $\Support{l-1}$, where $(\Support{l-1})_{ij} = \text{sign} [(\D{l-1})_{ij}]$.
    \item Lastly, $\Q{l-1} = I - \V{l-1}\Support{l-1}\V{l-1}^{\dagger}$ ensures that the pair of binary Kraus operators generates a valid POVM: $\sum_{\{0,1\}}\A{l}^{\dagger}\A{l} = I$.
\end{enumerate}
This construction ensures that the matrix~\ref{eq:stinespring_matrix} is unitary and that the cumulative Kraus operator 
\begin{equation}
    \A{L}^{c} = \A{L} ... \A{2}\A{1}
\end{equation}
corresponds to the correct POVM element: $\A{L}^{c \dagger}\A{L}^{c} = F_i$ with $i = b^{(L)}+1$, as shown in the Appendix to paper by Shen et al. ~\cite{Shen2017}.

Each time, the coupling unitary applied will depend on the sequence of previous measurement outcomes that define the current branch. Therefore, the scheme requires mid-circuit measurements and feed-forward. It is worth noting that the system is never measured directly during the binary search. This fact makes this approach also applicable to quantum channels, as shown by Shen et al.~\cite{Shen2017}.

In contrast to Naimark's dilation, which involves a single $M$-dimensional unitary applied to a set of $\mathrm{log}_2M$ qubits, the binary scheme requires a total of $\mathrm{log}_2M$ unitary operations, each acting on $n+1$ qubits.

\section{Extended noise analysis}\label{sec:detailed_noise_analysis}
In Section~\ref{sec:noise_analysis} of the main text, we employ a simple phenomenological noise model to explain the experimental results of the two-qubit experiment. The two hyperparameters of this model, $\epsilon_{\mathrm{CNOT}}$ and $\epsilon_{\mathrm{idle}}$, represent the depolarizing errors associated with each CNOT gate and each measurement and feedforward operation, respectively. While this model does not account for complex noise processes such as correlated errors, leakage, $T_1$ fluctuations, or measurement-induced control errors, which may occur on real superconducting qubit platforms~\cite{LukeGovia2022MCM}, the close agreement with the experimental results in Fig.~\ref{fig:detector_tomography_combined_figure} suggests that it effectively captures the main error contributions from noisy CNOT gates and noise related to mid-circuit measurements and idle time. 

\begin{figure*}[!tb]
    \centering
    \includegraphics{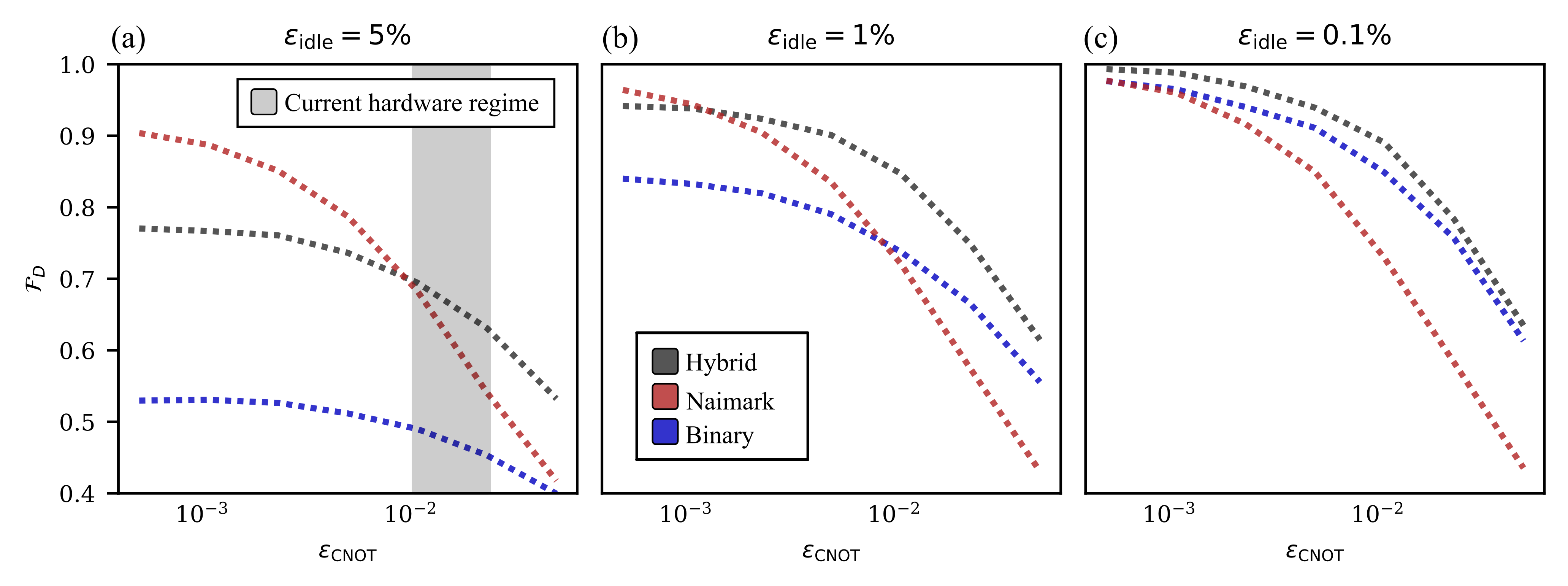}
    \caption{Noisy simulation of the two-qubit POVM for the binary tree (blue), Naimark's dilation (red), and hybrid (grey) schemes under various noise regimes. For each scheme, and for each combination of $\epsilon_{\mathrm{idle}}$ and $\epsilon_{\mathrm{CNOT}}$, the fidelity is computed for 10 circuits at varying CNOT depths and the highest fidelity value is plotted. The grey-shaded area indicates the regime where the simulations best agree with the experimental results obtained on hardware.}
    \label{fig:idle_cnot_error_sweep}
\end{figure*}

Here, we apply the same model to qualitatively understand the different noise regimes affecting the performance of our hybrid approach. Fig.~\ref{fig:idle_cnot_error_sweep} displays noisy simulations of the two-qubit SIC-POVM across various error regimes. For each combination of $\epsilon_{\mathrm{idle}}$ and $\epsilon_{\mathrm{CNOT}}$, we perform simulations for all three schemes at different CNOT depths, selecting for each scheme the depth that yields the highest fidelity. These fidelity levels are depicted in Fig.~\ref{fig:idle_cnot_error_sweep}.

Fig.~\ref{fig:idle_cnot_error_sweep} (a) represents the regime with comparatively high $\epsilon_{\mathrm{idle}}$, where the binary scheme is most affected due to a large number of conditional operations and mid-circuit measurements. In the range $\epsilon_{\mathrm{CNOT}} \approx 1-2\%$, we observe the best agreement with our experimental results. We thus conclude that the grey-shaded area in Fig.~\ref{fig:idle_cnot_error_sweep} (a) corresponds to the current hardware regime. Figs.~\ref{fig:idle_cnot_error_sweep} (b) and (c) show lower $\epsilon_{\mathrm{idle}}$ regimes, potentially achievable with faster mid-circuit measurements and conditional operations or through effective error-suppression techniques such as dynamical decoupling~\cite{bäumer2024quantum}. In these conditions, we expect the hybrid scheme to offer a more significant advantage over Naimark. Importantly, in all scenarios, the fidelity of Naimark decays fastest with increasing $\epsilon_{\mathrm{CNOT}}$ due to its higher CNOT depth.

Overall, while the fidelity improvements offered by the hybrid approach depend on noise from conditional operations and mid-circuit measurements, the exponential growth of CNOT depth with the system size is more detrimental compared to the linear scaling of mid-circuit measurements, as detailed in Section~\ref{sec:scaling_to_larger_systems}. Thus, while it is possible that the hybrid scheme might not offer an advantage for smaller system sizes on some experimental platforms with, e.g., high-quality CNOTs and poor-quality conditional operations, we expect hybrid to outperform its constituents on larger systems. In particular, we demonstrate that the hybrid approach already provides an advantage in a two-qubit system for IBM quantum devices.

\section{Data acquisition}\label{sec:data_acquisition}
We conducted both one- and two-qubit experiments on \texttt{ibmq\_kolkata}, a 27-qubit quantum processor~\cite{IBMQuantumPlatform}. For the one-qubit experiment, exact compiling was utilized. For the two-qubit experiment, we applied approximate compiling across 10 different CNOT depths for all three methods, yielding different levels of approximation accuracy. Subsequently, for each circuit, we generated five twirled instances by inserting random Pauli gates before and after CNOT gates. For each twirled instance, we prepared 36 two-qubit Pauli basis states and took 4,000 measurement samples for each state and instance, totaling 20,000 samples per circuit per basis state. The collected measurement statistics were then used to reconstruct the POVMs as outlined in Appendix~\ref{sec:detector_tomography_appendix}. From the reconstructed POVMs, we computed point estimates for the fidelities of each circuit. To estimate confidence intervals for the fidelity values, we used bootstrapping, resampling the measurement statistics to obtain a set of estimates from which we calculated the standard deviation. We executed 300 bootstrap instances for each circuit depth. The resulting confidence intervals, typically ranging from 0.1\% to 0.2\%, were too narrow to be visible on the plots in the main text. Overall, the experimental data was collected over the span of 2 hours, well within the typical noise drift timescale. Nevertheless, to mitigate any potential biases from the slowly drifting noise environment, we randomized the order of all circuits before submission to the quantum backend.

\bibliography{bibliography}

\end{document}